\documentclass[aps,prd,eqsecnum,amssymb,nofootinbib,floatfix,twocolumn,showpacs,a4paper]{revtex4-1}

\usepackage{amsmath}
\usepackage{mathtools}
\usepackage{dcolumn}
\usepackage{graphicx}
\usepackage{multirow}

\allowdisplaybreaks

\newcommand{\bdm}{\begin{displaymath}}
\newcommand{\edm}{\end{displaymath}}
\newcommand{\be}{\begin{equation}}
\newcommand{\ee}{\end{equation}}
\newcommand{\bse}{\begin{subequations}}
\newcommand{\ese}{\end{subequations}}
\newcommand{\ba}{\begin{array}}
\newcommand{\ea}{\end{array}}

\newcommand{\btau}{\boldsymbol{\tau}}
\newcommand{\btheta}{\boldsymbol{\theta}}
\newcommand{\bxi}{\boldsymbol{\xi}}

\newcommand{\Ca}{C_\textrm{a}}

\newcommand{\chii}{\chi_\text{i}}
\newcommand{\cmin}{C_\textrm{min}}

\newcommand{\F}{\mathcal{F}}

\newcommand{\fc}{f_\text{c}}

\newcommand{\md}{{\mathrm{d}}}

\newcommand{\mi}{{\mathrm{i}}}

\newcommand{\To}{T_\text{o}}

\newcommand{\ti}{t_\text{i}}

\newcommand{\dlangle}{\langle\langle}
\newcommand{\drangle}{\rangle\rangle}

\begin{document}

\title{Banks of templates
for all-sky narrow-band searches of gravitational waves
from~spinning~neutron~stars}

\author{Andrzej Pisarski}
\email{a.pisarski@uwb.edu.pl}
\author{Piotr Jaranowski}
\email{p.jaranowski@uwb.edu.pl}
\affiliation{Faculty of Physics,
University of Bia{\l}ystok,
K.\ Cio{\l}kowskiego 1L, 15--245 Bia{\l}ystok, Poland}


\date{\today}

\begin{abstract}

We construct efficient banks of templates suitable
for all-sky narrow-band searches
of almost monochromatic gravitational waves
originating from spinning neutron stars in our Galaxy
in data collected by interferometric detectors.
We consider waves with one spindown parameter included and
we assume that both the position of the gravitational-wave source in the sky
and the wave's frequency together with spindown parameter are unknown.
In the construction we employ simplified model of the signal with constant amplitude
and phase which is a linear function of unknown parameters.
Our template banks enable usage of the fast Fourier transform algorithm
in the computation of the maximum-likelihood $\F$-statistic for nodes of the grids defining the bank
and fulfill an additional constraint needed to resample the data to barycentric time efficiently.
All these template bank features were employed
in the recent all-sky $\F$-statistic-based search for continuous gravitational waves in Virgo VSR1 data
[J.\ Aasi \textit{et al.}, Classical Quantum Gravity \textbf{31}, 165014 (2014)].
Here we improve that template bank by constructing templates suitable for larger range of search parameters
and of smaller thicknesses for certain values of search parameters.
One of our template banks has thickness 12\% smaller than the thickness of the template bank
used in the all-sky search of Virgo VSR1 data
and only 4\% larger than the thickness of 4-dimensional optimal lattice covering $A_4^\star$.

\end{abstract}

\pacs{95.55.Ym, 04.80.Nn, 95.75.Pq, 97.60.Gb}

\maketitle

\section{Introduction}

The era of first-generation ground-based interferometric gravitational-wave detectors
(LIGO \cite{LIGO}, Virgo \cite{Virgo}, GEO600 \cite{GEO600}, and TAMA300 \cite{TAMA300})
is over and the detectors are now undergoing major upgrades.
One of the primary sources of gravitational waves
for both first-generation detectors as well as their advanced, second-generation versions
(including advanced LIGO \cite{aLIGO} and advanced Virgo \cite{aVirgo} instruments),
are rotating neutron stars located in our Galaxy
(\cite{CutlerThorne2002}; see Ref.\ \cite{2012--Weinstein}
for short review of astronomy and astrophysics with gravitational waves in the advanced detector era).
They are expected sources of \emph{almost monochromatic} gravitational waves
(which are often a bit too broadly called just continuous waves)
and in the present paper we consider the specific problem
related with the \emph{maximum-likelihood $\F$-statistic-based detection}
of this kind of waves in the detector's noise: construction of efficient banks of templates.

Depending on what is \emph{a priori} known about sources
of gravitational waves we are looking for,
searches can be splitted into targeted, directed, and all-sky (or blind) searches.
In \emph{targeted} searches both the position of the source in the sky
and the wave's frequency together with spindown parameters
(i.e., the time derivatives of the frequency evaluated at some reference moment of time) are known.
If one assumes that only the position of the source in the sky is known
but one does not know the frequency and the spindown parameters,
one performs \emph{directed} searches for gravitational-wave signals.
Finally, in \textit{all-sky} or \textit{blind} searches one assumes
that both the position of the source in the sky
and the frequency and spindown parameters are not known.

Several searches for almost monochromatic gravitational waves
originating from spinning neutron stars in our Galaxy
were already performed in the data collected by
the first-generation LIGO, Virgo, and GEO600 detectors.
The results of \emph{targeted} searches were published in Refs.\
\cite{LSC04,LSC05,LSC07,LSC08,LSC10,LSC.VC...2011,2014--LSC-VC--ApJ,2014--LSC-VC--astroph}.
Among them searches for gravitational waves from Crab and Vela pulsars were
reported in \cite{LSC08,LSC10} and \cite{LSC.VC...2011} (see also \cite{2012--Gill}), respectively,
and a more recent targeted search (using data collected during
LIGO science run S6 and Virgo runs VSR2, VSR4) was presented in Ref.\ \cite{2014--LSC-VC--ApJ}
(together with the most up-to-date results from all targeted pulsar searches
performed on data collected by the first-generation detectors).
The results of the first \emph{directed} searches for gravitational waves
from solitary neutron stars were published in Refs.\ \cite{LSC.VC...2010,2013--LSC-VC--PRD}:
from the supernova remnant Cassiopeia A in \cite{LSC.VC...2010}
and from the Galactic center in \cite{2013--LSC-VC--PRD}.
Another kind of directed search is search for gravitational waves
from sources in known binary systems but with unknown frequency.
The results of search of this kind from the brightest low-mass X-ray binary Scorpius X-1
were recently reported in \cite{2014--LSC-VC--grqc}.
Results of \emph{all-sky} searches of data collected during
LIGO science runs S2--S5 were reported in Refs.\
\cite{LSC...2005,LSC...2007,LSC...2008,LSC...2009,LSC.VC...2012,2014--LSC-VC--CQG,EH...2009a,EH...2009b,2013--EH}.
They include all-sky searches on S4 and S5 LIGO data performed within an Einstein@Home initiative
\cite{EH...2009a,EH...2009b,2013--EH} (running on the BOINC infrastructure \cite{BOINC}).
An all-sky search in Virgo VSR1 data was presented in \cite{2014--LSC-VC--CQG-b}.
Let us finally also mention the report from the first all-sky search
for continuous waves from unknown sources located in binary systems \cite{2014--LSV-VC--PRD}.
The results of the searches for continuous waves
with the first-generation LIGO and Virgo detectors
were shortly reviewed in \cite{2012--Astone,2012--Krolak}.

In all these searches several different data analysis strategies were employed.
We expect that the gravitational-wave signal coming from a rotating neutron star
is so week that to detect it in the detector's noise
one has to analyze months-long segments of data.
Fully coherent analysis of such amount of data is computationally prohibitive
in the case of all-sky searches \cite{BCCS98,BC99},
therefore different hierarchical two-stage schemes were developed,
where in the first stage shorter segments of data are analyzed coherently
and then in the second stage the results are combined in an incoherent way.

In the present paper we consider only coherent detection
of almost monochromatic gravitational-wave signals
which can be employed in the first stage of a hierarchical two-stage procedure.
Moreover we restrict ourselves to detection based on the \textit{maximum-likelihood} (ML) principle
which leads to the detection statistic
(by means of which one can test whether data contains gravitational-wave signal)
known as the \emph{$\F$-statistic} (originally introduced in \cite{JKS98}).
We also assume that the noise in the detector is \emph{Gaussian} and \emph{stationary}
(details of the ML detection in Gaussian noise can be found
e.g.\ in \cite{W71,JKlrr} and in Chapter 6 of \cite{JKbook}).
Finally we consider \emph{narrow-band} searches
(at the beginning of Sec.\ II we explain what does it precisely mean).
Data analysis tools and algorithms needed to perform,
within the ML approach, an all-sky narrow-band search
for almost monochromatic gravitational-wave signals
were developed in detail in the series
of papers \cite{JKS98,JK99,JK00,ABJK02,ABJPK10}
(see also Refs.\ \cite{BCCS98,BC99}).

The detection $\F$-statistic (together with its modifications) was employed in several searches
for almost monochromatic gravitational waves performed so far.
Different detection statistics needed to perform targeted searches
were derived in \cite{Dupuis-Woan-2005} and then in Refs.\ \cite{PK09,JK10} (from the ML principle
and also using the Bayesian approach together with the composite hypothesis testing)
and were used (among other approaches) e.g.\ in searches reported in \cite{LSC.VC...2011}.
The $\F$-statistic was used in the directed search reported in Refs.\ \cite{LSC.VC...2010}
(see also \cite{Wette.et.al...2008}), \cite{2014--LSC-VC--grqc},
and in the first stage of Einstein@Home all-sky searches of \cite{EH...2009a,EH...2009b}.
The results of an all-sky search reported in Ref.\ \cite{2014--LSC-VC--CQG-b}
were achieved by means of an implementation of the $\F$-statistic based pipeline
and in the current work we improve bank of templates used in that search.
The $\F$-statistic was recently generalized in Ref.\ \cite{Keitel-others-2014} 
by adding an explicit simple line hypothesis to Gaussian noise hypothesis underlying the standard $\F$-statistic.

The ML detection of the gravitational-wave signal with unknown parameters $\btheta$
relies on maximization of the likelihood ratio $\Lambda[x;\btheta]$
(which depends on the data $x$) with respect to the parameters $\btheta$
and comparing this maximum with a threshold.
In the case of directed or all-sky searches the parameters $\btheta$
form two groups, $\btheta=(\mathbf{A},\bxi)$.
There are four \emph{extrinsic} (sometimes called, not very precisely, \emph{amplitude}) parameters $\mathbf{A}$:
an overall amplitude and initial phase of the waveform,
the polarization angle of the wave and the inclination angle
(of the star's rotation axis with respect to the line of sight).
\emph{Intrinsic} (also called \emph{phase}) parameters $\bxi$ consist of
the frequency of the wave, the spindown parameters,
and the two more parameters depending on the position of the gravitational-wave source in the sky
(they are known in the case of directed searches).
Maximization of the $\Lambda$ with respect to parameters $\mathbf{A}$ is done analytically
by solving equations $\partial\Lambda/\partial\mathbf{A}=0$;
their solution with respect to $\mathbf{A}$ defines the ML estimators
$\hat{\mathbf{A}}=\hat{\mathbf{A}}[x;\bxi]$
of the parameters $\mathbf{A}$.
Then the \textit{$\F$-statistic} is defined as the logarithm of the likelihood ratio $\Lambda$
in which the amplitude parameters $\mathbf{A}$ are replaced
by their ML estimators $\hat{\mathbf{A}}$:
$\F[x;\bxi]:=\ln\Lambda[x;\hat{\mathbf{A}}[x;\bxi],\bxi]$.

Maximization of the $\F$-statistic over the phase parameters $\bxi$ can be done only numerically.
To find the maximum of the $\F$-statistic
one constructs a \textit{bank of templates} in the space of the parameters $\bxi$,
which is determined by a discrete set of points, i.e.\ a \emph{grid} in the parameter space.
The grid is chosen in such a way, that for any possible gravitational-wave signal present in data
there exists a grid point such that the relative loss
(with respect to the value achieved for exact matching between the parameters of the signal
and one of the grid points) of the expectation value of the $\F$-statistic
computed for the parameters of this grid point is not less than a certain fixed minimal value.

In the case of directed searches in the construction of the template banks
one can use the simplified \textit{polynomial phase model} of the gravitational-wave signal.
This model was first introduced in Ref.\ \cite{BCCS98}
and then more in detail in \cite{JK99} (see Sec.\ V D and Appendix C there);
in the model the signal's amplitude is constant and the phase is a polynomial function of time.
It was found (in Sec.\ V E of \cite{JK99}) that in the case of directed searches
the polynomial phase model reproduces very well the covariance matrix
(defined as the inverse of the Fisher matrix)
for the ML estimators of the parameters of the exact gravitational-wave signal.
The polynomial model was used in the search reported in Ref.\ \cite{LSC.VC...2010}
(see also \cite{Wette.et.al...2008}). 
In this search the phase of the gravitational-wave
signal was modeled as a third-order-in-time polynomial
(i.e.\ the frequency of the wave and its first and second spindown
parameters were taken into account),
and a template bank based on a body-centered cubic lattice was used.
Efficient banks of templates for a second-order-in-time polynomial phase model
was recently constructed in Ref.\ \cite{PJP2011}.

In the present paper we are interested in all-sky searches,
in which another simplified model of the gravitational-wave signal
can be employed in the construction of the template banks.
This is so-called \textit{linear phase model},
in which the amplitude of the signal is constant
and the signal's phase is a linear function of the unknown parameters.
The model was introduced in Sec.\ V B of Ref.\ \cite{JK99},
and in Sec.\ V E of \cite{JK99} it was shown that the model reproduces well
the covariance matrix (defined again as the inverse of the Fisher matrix)
for the ML estimators of the parameters of the exact gravitational-wave signal.
This model is used in the present paper.

For the linear phase model the expectation value of the $\F$-statistic depends
[see the key Eq.\ \eqref{EF3} below] on the signal-to-noise ratio $\rho$
and on the value of the \textit{autocovariance function} $C_0(\bxi,\bxi')$\footnote{
In our considerations the autocovariance function $C_0$ plays the role of a \emph{match},
which is the notion commonly used in the literature on template placing
based on the concept of a metric in the space spanned by parameters of the signal
(see, e.g., \cite{BSD1996,Owen1996} and Appendix \ref{Ametric}).} of the $\F$-statistic
(the subscript `0' indicates that the autocovariance is calculated in the case when data is a pure noise)
computed for the intrinsic parameters of the template ($\bxi$)
and the gravitational-wave signal ($\bxi'$), respectively.
The signal-to-noise ratio $\rho$ we can not control, therefore we fix its minimal value.
Then to construct bank of templates
one needs to choose some minimum value $\cmin$\footnote{
$\cmin$ can be identified with the \emph{minimal match} used in the ``metric'' approach
to the problem of template placement; see Appendix \ref{Ametric}.}
of the autocovariance function $C_0$
and look for such a grid of points that for any point $\bxi'$ in the intrinsic parameter space
there exists a grid node $\bxi$ such that the autocovariance $C_0(\bxi,\bxi')$ computed for the parameters
$\bxi$ and $\bxi'$ is not less than $\cmin$. The autocovariance $C_0(\btau)$
(for linear phase model it depends on $\bxi$, $\bxi'$ only through the difference $\btau:=\bxi-\bxi'$)
can be approximated by taking the Taylor expansion (up to the second-order terms) of $C_0$
around its maximum at $\btau=\mathbf{0}$.
Then isoheights of such approximated autocovariance function $C_0$ are \emph{hyperellipsoids}.
The problem of constructing bank of templates can thus be formulated
as a problem of finding \emph{optimal covering} of the signal's parameter space by means of
identical hyperellipsoids defined as isoheights of the autocovariance function $C_0$ of the $\F$-statistic.

In our paper we are interested in such searches for almost monochromatic gravitational-wave signals
for which the number of grid points in the parameter space is very large
and the time needed to compute the $\F$-statistic for all grid nodes is long.
Then it is crucial to use in the computation the fast numerical algorithms.
Because the computation of the $\F$-statistic involves calculation of the Fourier transform,
one would like to use the \emph{fast Fourier transform} (FFT) algorithm.
In an all-sky search described in Ref.\ \cite{2014--LSC-VC--CQG-b}
the FFT algorithm was used in the coherent part of the search,
what resulted in a fifty-fold speed up in computation of the $\F$-statistic
compared to the algorithms used in other data-analysis procedures.
The FFT algorithm computes the values of the \emph{discrete Fourier transform} (DFT)
of a time series for a certain set of discrete frequencies called the \emph{Fourier frequencies}.
Thus it will be possible to use the FFT algorithm in computation of the $\F$-statistic,
provided the grid points will be arranged in such a way,
that the frequency coordinates of these points
will all coincide with the Fourier frequencies.
All grids constructed in our paper fulfill this requirement.

The construction of efficient banks of templates for matched-filtering searches
was considered in Ref.\ \cite{P07}. The usage of random template banks and relaxed lattice coverings
for gravitational-wave searches was recently discussed in Ref.\ \cite{MPP09}
(see also \cite{HAS09,MV10,R10}).
More recently an efficient lattice template placement for coherent all-sky searches
based on a new flat parameter-space metric
was proposed in Refs.\ \cite{2013--Wette-Prix,2014--Wette}.
As explained above in the current paper we are interested in searches involving
data streams so long, that the time performance of the search crucially
depends on the ability of using the FFT algorithm.
This enforces the above-mentioned constraint
which is not always, i.e., not for all grids and/or not for all values of search parameters,
fulfilled by grids considered in Refs.\ \cite{P07,MPP09,HAS09,MV10,R10,2013--Wette-Prix,2014--Wette}.
Therefore our work can be considered as being complementary to the studies performed
in Refs.\ \cite{P07,MPP09,HAS09,MV10,R10,2013--Wette-Prix,2014--Wette}.
The grids constructed in our paper enable free choice
of search parameters, i.e., the number $N$ of data points to be analyzed,
the number $N_{\text{FFT}}$ of data points in the time series being Fourier transformed
(which is different from the previous one when zero padding is employed, see Sec.\ III A below),
and the quantity $\cmin$ introduced above. This flexibility is important
because to speed up computation of the $\F$-statistic for all grid nodes
one should ensure that $N_{\text{FFT}}=2^n$ with $n$ being positive integer
(then the FFT algorithm is the fastest), and one can also choose the 
observational interval to be integer multiple of one sidereal day,
then the analytic formula for the $\F$-statistic considerably simplifies.\footnote{
The simple formula for the $\F$-statistic used in the search of Ref.\ \cite{2014--LSC-VC--CQG-b}
and given in Eq.\ (9) there is valid only if the observational interval is an integer multiple of one sidereal day.}
One can however start from the best known lattice for the given number of the unknown parameters\footnote{
We thank the anonymous referee for pointing out this possibility.}
(e.g.\ the lattice $A_4^\star$ in the case of four parameters) and then manipulate the numerical values
of the search parameters $N$, $N_{\text{FFT}}$, and $\cmin$ in such a way
that the frequency coordinates of points of this best known lattice
will all coincide with the Fourier frequencies.

Grids enabling the use of the FFT algorithm in the ML detection of gravitational-wave signals
from white-dwarf binaries in the mock LISA data challenge were devised in Ref.\ \cite{BBK2010}
(see also \cite{BKP2009,BKB2009}),
where the geometric approach
(initialized in \cite{BSD1996,Owen1996} for searches of gravitational waves
from inspiralling compact binaries
and then developed also for searches of continuous gravitational waves,
see \cite{Prix2007}) was employed
and the grids were constructed by some deformation of the optimal lattice coverings
$A_d^\star$ in $d=3,4$ dimensions.
The algorithm needed to construct templates
for all-sky narrow-band searches for almost monochromatic gravitational waves
fulfilling the FFT-related constraint was devised in Sec.\ IV of Ref.\ \cite{ABJPK10}.
The bank of templates generated by means of this algorithm
was used in the recent all-sky search of Virgo VSR1 data reported in \cite{2014--LSC-VC--CQG-b}.
In Sec.\ V below we compare the grids constructed in the present paper
with those developed in \cite{ABJPK10}.

The organization of the paper is as follows.
In Sec.\ II we introduce the linear phase model of the gravitational-wave signal.
We consider here the phase with one spindown parameter included.
For this model we compute the $\F$-statistic and its expectation value
in the case when the data contains the gravitational-wave signal.
In Sec.\ III we introduce some mathematical notions related with coverings
and we formulate the constraints we want to force on grids.
Section IV is devoted to construction of two different families of grids.
The grids enable usage of the FFT algorithm
in the computation of the $\F$-statistic for nodes of the grids
and fulfill an additional constraint needed to resample the data to barycentric time efficiently.
In Sec.\ V we discuss our results.
In Appendix \ref{Ametric} we compare the language employed by us in the present paper
(based on the notion of autocovariance function of the $\F$-statistic)
with the language of a~metric in the space of signal's parameters.
Appendix \ref{ALengths} gives some details of the optimal 4-dimensional $A_4^\star$ lattice
and Appendix \ref{ACovering} contains a brief sketch of the algorithm
we use to find covering radius of given lattice.

\section{Autocovariance function of the $\F$-statistic}

We assume that the noise $n$ in the detector is an additive, stationary, Gaussian,
and zero-mean continuous stochastic process.
Then the logarithm of the likelihood function is given by
\be
\label{001}
\ln\,\Lambda[x] = (x|h)-\frac{1}{2}\,(h|h),
\ee
where $x$ denotes the data from the detector,
$h$ is the deterministic signal we are looking for in the data,
and $(\cdot\,|\,\cdot\,)$ is the scalar product between waveforms defined by
\be
\label{001a}
(h_{1}|h_{2}) := 4\,\text{Re}
\int_{0}^{\infty} \frac{\tilde{h}_1(f)\,{\tilde{h}}_2^*(f)}{S_n(f)}\:\md f.
\ee
Here \hspace*{1ex}$\tilde{}$\hspace*{1ex} stands for the Fourier transform,
* denotes complex conjugation, and $S_{n}$ is the \textit{one-sided} spectral density
of the detector's noise $n$ ($S_n$ is defined thus for frequencies $0\le f<+\infty$).

We are interested in narrow-band searches for \textit{almost monochromatic} signals,
i.e.\ such signals for which the modulus of the Fourier transform is well concentrated
(for frequencies $f\ge0$) around some fixed frequency.
The search is \emph{narrow-band} in the sense that the frequency bandwith of the search
is small enough to assume that the spectral density $S_{n}$
is a slowly changing function of $f$ within the bandwith.
Then, if both waveforms $h_1$ and $h_2$ in Eq.\ \eqref{001a} have their Fourier transforms
concentrated around frequencies within the bandwith of the search,
we can replace $S_{n}(f)$ in the integrand of \eqref{001a} by a constant $S_{n}(\fc)$,
where $\fc$ is some `central' frequency of the bandwith.
Consequently, after employing the Parseval's theorem,
we approximate the scalar product \eqref{001a} by
\be
\label{002}
(h_{1}|h_{2}) \cong \frac{2}{S_{n}(\fc)}\,
\int\limits_{\ti-\To/2}^{\ti+\To/2}h_{1}(t)\,h_{2}(t)\:\md t
= \frac{2\,\To}{S_{n}(\fc)}\langle h_{1}h_{2}\rangle.
\ee
Here $\langle\ti-\To/2;\,\ti+\To/2\rangle$ denotes observational interval,
so $\To$ is the length of observation time,
and $\ti-\To/2$ is the moment at which the observation begins.
The time averaging operator $\langle\,\cdot\,\rangle$ is defined by
\be
\label{003}
\langle h \rangle := \frac{1}{\To}
\int\limits_{\ti-\To/2}^{\ti+\To/2}h(t)\,\md t.
\ee
Using the formula \eqref{002} we can write the log likelihood ratio from Eq.\ \eqref{001} as
\be
\label{004}
\ln\Lambda[x] \cong \frac{2\,\To}{S_{n}(\fc)}\,
\Big(\langle x\,h \rangle -\frac{1}{2}\langle h^{2}
\rangle\Big).
\ee

In construction of template banks we employ an approximate model
of the continuous gravitational-wave signal from a rotating neutron star
(this model was introduced in Sec.\ V B of Ref.\ \cite{JK99},
where it was called ``linear model I''; it should be distinguished from ``linear model II''
introduced in Sec.\ V C of \cite{JK99} and not used in the current paper).
The approximation relies on
(i) assuming that the amplitude of the signal is constant,
so we neglect the slowly varying modulation of the signal's amplitude
due to motion of the detector with respect to the
solar system barycenter (SSB);
(ii) neglecting these terms in the phase modulation
due to motion of the detector with respect to the SSB
which depend on spin downs;
(iii) discarding perpendicular to the ecliptic component of the vector
connecting the SSB and the detector.
This leads to the signal's model which is called {\em linear}
because it has the property that its phase
is a linear function of the parameters.
The approximate signal's model can be written as
\be
\label{h}
h(t;h_0,\Phi_0,\bxi)
= h_{0}\,\sin\big(\Phi(t;\bxi)+\Phi_{0}\big),
\ee
where $h_0$ ia a constant amplitude and $\Phi_0$ is a constant initial phase.
The time-dependent part $\Phi(t;\bxi)$ of the phase depends on the $s+3$ parameters $\bxi$,
\be
\bxi = (\omega_0,\ldots,\omega_s,\alpha_1,\alpha_2),
\ee
and has the following form
\begin{align}
\label{Phi}
\Phi(t;\bxi) = \sum_{k=0}^{s} \omega_k \Big(\frac{t}{\To}\Big)^{k+1}
+ \alpha_{1}\,\mu_{1}(t) + \alpha_{2}\,\mu_{2}(t).
\end{align}
The dimensioneless parameters $\omega_k$ are defined as
\be
\label{omegak}
\omega_k := \frac{2\pi}{(k+1)!} f_0^{(k)} \To^k,
\quad k=0,\ldots,s,
\ee
where $f_{0}^{(0)}\equiv f_0$ is an instantaneous frequency of the gravitational wave
computed at the SSB at $t=0$ and $f_{0}^{(k)}$ ($k=1,\ldots,s$)
is the $k$th time derivative
of the instantaneous gravitational-wave frequency at the SSB evaluated at $t=0$.
The parameters $\alpha_1$ and $\alpha_2$ are related with the position of the gravitational-wave source
in the sky through the definitions
\bse
\begin{align}
\alpha_1 &:= 2\pi f_{0}(\sin\alpha\cos\delta\cos\varepsilon+\sin\delta\sin\varepsilon),
\\[2eX]
\alpha_2 &:= 2\pi f_{0}\cos\alpha\cos\delta,
\end{align}
\ese
where $\alpha$ is the right ascension and $\delta$ is the declination of the source,
$\varepsilon$ is the obliquity of the ecliptic.
The functions $\mu_1(t)$ and $\mu_2(t)$ are known functions of time,
\bse
\label{mu1mu2}
\begin{align}
\mu_{1}(t):=&\frac{1}{c}(R_{\text{ES}}^{y}(t)+R_{\text{E}}^{y'}(t)\cos\varepsilon),
\\[2eX]
\mu_{2}(t):=&\frac{1}{c}(R_{\text{ES}}^{x}(t)+R_{\text{E}}^{x'}(t)),
\end{align}
\ese
where $(R^x_\mathrm{ES},R^y_\mathrm{ES},0)$
are the components of the vector joining the SSB
with the center of the Earth in the SSB coordinate system,
and $(R^{x'}_\text{E},R^{y'}_\mathrm{E},R^{z'}_\mathrm{E})$ are the
components of the vector joining the center of the Earth
and the detector's location in the celestial coordinate system.\footnote{
The definitions of the SSB and celestial coordinate systems
are given in Sec.\ II of Ref.\ \cite{JKS98}.}

Let us introduce two new parameters,
\be
\label{21}
h_{1}:=h_{0}\cos{\Phi_{0}},
\quad
h_{2}:=h_{0}\sin{\Phi_{0}}.
\ee
Then $h_0=\sqrt{h_1^2+h_2^2}$ and the signal $h$ can be written as follows
\be
\label{22}
h(t;h_1,h_2,\bxi) = h_{1}\sin\Phi(t;\bxi) + h_{2}\cos\Phi(t;\bxi).
\ee
The time average $\langle{h^2}\rangle$ equals
\be
\label{23}
\langle h^{2} \rangle = \frac{1}{2}(h_{1}^{2} + h_{2}^{2}) +
\frac{1}{2}(h_{2}^{2}-h_{1}^{2}) \langle\cos2\Phi\rangle
+ h_{1}h_{2} \langle \sin2\Phi \rangle.
\ee
For observations which last at least several hours
and for gravitational-wave frequencies of the order of tens of Hertz or higher,
to a good approximation
\be
\label{24}
\langle\sin2\Phi\rangle\cong0,
\quad
\langle\cos2\Phi\rangle\cong0.
\ee
Then the time average \eqref{23} simplifies to
\be
\label{25}
\langle h^{2} \rangle \cong \frac{1}{2}(h_{1}^{2} + h_{2}^{2}).
\ee
Making use of \eqref{25} we compute the optimal signal-to-noise ratio
$\rho$ for the signal \eqref{22}:
\be
\label{snr}
\rho(h_0) = \sqrt{(h|h)}
\cong \sqrt{\frac{2\To}{S_{n}(\fc)} \langle h^{2} \rangle}
\cong h_0 \sqrt{\frac{\To}{S_{n}(\fc)}}.
\ee

Substituting Eqs.\ \eqref{22} and \eqref{25} into \eqref{004}
we get the following formula for the log likelihood ratio of the signal \eqref{22}:
\begin{align}
\label{26}
\ln\Lambda[x;h_{1},h_{2},\bxi] &\cong
\frac{2\,\To}{S_{n}(\fc)}\,\bigg(h_{1}\,\langle x(t)\sin\Phi(t;\bxi)\rangle
\nonumber\\
&\quad + h_{2}\,\langle x(t)\,\cos\Phi(t;\bxi)\rangle-\frac{1}{4}(h_{1}^{2}+h_{2}^{2})\bigg).
\end{align}
Next we maximize $\ln\Lambda$ with respect to the parameters $h_{1}$ and $h_{2}$
by solving equations
\be
\label{27}
\frac{\partial\ln\Lambda[x;h_{1},h_{2},\bxi]}{\partial h_{i}}=0,\quad i=1,2.
\ee
The unique solution to these equations,
\be
\label{28}
\widehat{h}_{1}\cong2\langle x\sin\Phi\rangle,
\quad
\widehat{h}_{2}\cong2\langle x\cos\Phi\rangle,
\ee
gives the maximum-likelihood estimators of the parameters $h_{1}$ and $h_{2}$.
After replacing in Eq.\ \eqref{26} the parameters $h_{1}$ and $h_{2}$
by their estimators $\widehat{h}_{1}$ and $\widehat{h}_{2}$,
we obtain the reduced log likelihood ratio which is called the \emph{$\F$-statistic}:
\begin{align}
\label{29}
\F[x;\bxi] &:= \ln \Lambda[x;\hat{h}_{1},\hat{h}_{2},\bxi]
\nonumber\\
&\cong \frac{2\,\To}{S_{n}(f_{c})}
\bigg(\langle x(t)\,\sin\Phi(t;\bxi)\rangle^{2}+\langle x(t)\,\cos\Phi(t;\bxi)\rangle^{2}\bigg).
\end{align}
Making use of $\exp(-\mi\Phi)=\cos\Phi-\mi\sin\Phi$ (for $\Phi\in\mathbb{R}$)
and the definition \eqref{003},
it is easy to rewrite the $\F$-statistic \eqref{29} in the form
\be
\label{30}
\F[x;\bxi] \cong \frac{2}{S_{n}(\fc)\To} \left|\,
\int\limits_{\ti-\To/2}^{\ti+\To/2} x(t) \exp\Big(-\mi \Phi(t;\bxi)\Big) \, \md t \, \right|^2.
\ee

We now study the expectation value of the $\F$-statistic \eqref{29} in the case
when the data $x$ contains some gravitational-wave signal $h$, i.e.
\be
x(t) = n(t) + h(t;\btheta'),
\ee
where $\btheta'=(h_1',h_2',\bxi')$ collects the parameters of the gravitational-wave
signal present in the data. We want thus to compute
\be
\label{EF1}
\text{E}_1\{\F[x;\bxi]\}
= \text{E}\big\{\F\big[n(t) + h(t;\btheta');\bxi\big]\big\},
\ee
where the subscript `1' means that the expectation value is computed in the case
when the data contains some signal.
One can show that
\begin{align}
\label{EF2}
\text{E}_1\{\F[x;\bxi]\}
&\cong 1 + \frac{1}{2} \, \rho(h_0')^2 \,
\Big( \big\langle\sin\big[\Phi(t;\bxi)-\Phi(t;\bxi')\big]\big\rangle^{2}
\nonumber\\
&\quad + \big\langle\cos\big[\Phi(t;\bxi)-\Phi(t;\bxi')\big]\big\rangle^{2} \Big),
\end{align}
where $\rho(h_0')$ is the signal-to-noise ratio from Eq.\ \eqref{snr}
computed for the signal $h(t;\btheta')$
(so $h_0'=\sqrt{h_1'^2+h_2'^2}$).
The right-hand side of the above equation can be expressed in terms
of the \emph{autocovariance function} $C_0$ of the $\F$-statistic 
(the subscript `0' means here that the autocovariance is computed in the case
when the data contains only noise).
In the signal-free case the $\F$-statistic $\mathcal{F}[n;\bxi]$
is the \emph{random field} which depends on the parameters $\bxi$,
and its autocovariance function is defined as
\be
\label{C01}
C_0(\bxi,\bxi') \coloneqq \mathrm{E}\big\{[\F[n;\bxi]-m_0(\bxi)]
[{\cal F}[n;\bxi']-m_0(\bxi')]\big\},
\ee
where $m_0$ is the signal-free expectation value of $\F$:
\be
\label{m0}
m_0(\bxi) \coloneqq \mathrm{E}\{\F[n;\bxi]\}.
\ee
In Sec.\ IV of Ref.\ \cite{JK00} it was shown that
the autocovariance function $C_0$ of the $\F$-statistic
for the narrow-band gravitational-wave signal of the form \eqref{22} 
can be approximated by
\begin{align}
\label{C02}
C_0(\bxi,\bxi') &\cong \big\langle\sin\big[\Phi(t;\bxi)-\Phi(t;\bxi')\big]\big\rangle^{2}
\nonumber\\
&\quad + \big\langle\cos\big[\Phi(t;\bxi)-\Phi(t;\bxi')\big]\big\rangle^{2},
\end{align}
therefore the expectation value \eqref{EF2}
can be written as
\be
\label{EF3}
\text{E}_1\{\F[x;\bxi]\}
\cong 1 + \frac{1}{2} \, \rho(h_0')^2 \, C_0(\bxi,\bxi').
\ee

The phase $\Phi$ of the gravitational-wave signal \eqref{22}
depends linearly on the parameters $\bxi$ [see Eq.\ \eqref{Phi}],
therefore the autocovariance \eqref{C02} 
depends only on the differences between $\bxi$ and $\bxi'$:
\be
\label{C03}
C_0(\bxi,\bxi') \cong \langle\sin\Phi(t;\bxi-\bxi')\rangle^{2}
+ \langle\cos\Phi(t;\bxi-\bxi')\rangle^{2}.
\ee
If one introduces $\btau:=\bxi-\bxi'$, one can thus write
\be
\label{C04}
C_0(\btau) \cong \langle\cos\Phi(t;\btau)\rangle^{2}
+ \langle\sin\Phi(t;\btau)\rangle^{2}.
\ee
Let us note that $C_0$ attains its maximal value equal to 1
for $\btau=\mathbf{0}$ (i.e.\ for $\bxi=\bxi'$).

We will further approximate the formula \eqref{C04} for the autocovariance function
in the case when $|\btau|\ll1$. We will also restrict ourselves
to the phase $\Phi$ of the signal depending on one spindown parameter,
so the phase is of the form [see Eq.\ \eqref{Phi}]
\be
\label{faza2}
\Phi(t;\btau) = \omega_{0}\frac{t}{\To} + \omega_{1}\bigg(\frac{t}{\To}\bigg)^2
+\alpha_{1}\,\mu_{1}(t)+\alpha_{2}\,\mu_{2}(t),
\ee
where the vector $\btau$ enclosing phase parameters has four components,
\be
\btau=(\omega_{0},\omega_{1},\alpha_{1},\alpha_{2}).
\ee
The approximation relies on expanding the right-hand side of Eq.\ \eqref{C04}
in Taylor series around $\btau=\mathbf{0}$ up to terms quadratic in $\btau$.
Such computed autocovariance function we denote by $\Ca$.
Making use of the obvious equalities
\be
\Phi(t;\btau=\mathbf{0}) = 0,
\quad
\frac{\partial^{2}\Phi}{\partial\tau_{k}\partial\tau_{l}}(t;\btau) = 0,
\quad k,l=1,\ldots,4,
\ee
we get
\be
\label{Ca}
C_0(\btau) \cong \Ca(\btau) \coloneqq
1 - \sum_{k=1}^4\sum_{l=1}^4 \tilde{\Gamma}_{kl}\,\tau_{k}\,\tau_{l},
\ee
where $\tilde{\Gamma}$ is the 4-dimensional
\emph{reduced Fisher matrix} with elements equal to
\be
\label{Gamma}
\tilde{\Gamma}_{kl} := \Big\langle\frac{\partial\Phi}{\partial\tau_{k}}
\frac{\partial\Phi}{\partial\tau_{l}}\Big\rangle -
\Big\langle\frac{\partial\Phi}{\partial\tau_{k}}
\Big\rangle \Big\langle\frac{\partial\Phi}{\partial\tau_{l}}
\Big\rangle, \quad k,l=1,\ldots,4.
\ee
Let us note that because the phase $\Phi(t;\btau)$ is a linear function of the parameters $\btau$,
the elements $\tilde{\Gamma}_{kl}$ of the Fisher matrix $\tilde{\Gamma}$ are constant:
they \emph{do not depend} on the values of the parameters $\btau$.

It is convenient to introduce the dimensionless quantity $\chii$
and to replace the time $t$ by the dimensionless variable $x$,
\be
\label{xchii}
\chii \coloneqq \frac{\ti}{\To},
\qquad
x \coloneqq \frac{t}{\To} - \chii.
\ee
The observational interval $\langle\ti-\To/2;\,\ti+\To/2\rangle$ of the time $t$
is transformed, according to \eqref{xchii},
into the interval $\langle-1/2;1/2\rangle$ of unit length.
Averaging with respect to the variable $x$ is thus defined as
\be
\langle\langle g(x) \rangle\rangle \coloneqq \int\limits_{-1/2}^{1/2}g(x)\,\md x.
\ee
It is easy to see that for any function of time $f(t)$ we have
\begin{align}
\langle f(t) \rangle
&= \frac{1}{\To}\int\limits_{\ti-\To/2}^{\ti+\To/2}f(t)\,\md t
\nonumber\\[1ex]
&= \int\limits_{-1/2}^{1/2}f(t(x))\,\md x
= \dlangle f(t(x)) \drangle,
\end{align}
where [see Eq.\ \eqref{xchii}] $t(x)=(x+\chii)\To$.
Making use of the above introduced definitions, the Fisher matrix $\tilde{\Gamma}$
with elements given in Eq.\ \eqref{Gamma}
and for the phase $\Phi$ defined in Eq.\ \eqref{faza2}
can be written as
\begin{widetext}
 \begin{align}
 \label{F5}
 \tilde{\Gamma}(\chii)
= \begin{pmatrix}
  \frac{1}{12}
& \frac{1}{6}\,\chii
& \dlangle x \mu_{1}\drangle
& \dlangle x \mu_{2}\drangle
\\[1ex]
  \frac{1}{6}\,\chii
& \frac{1}{180}+\frac{1}{3}\,\chii^{2}
& \tilde{\Gamma}_{23}
& \tilde{\Gamma}_{24}
\\[1ex]
  \dlangle x\mu_{1} \drangle
& \tilde{\Gamma}_{32}
& \dlangle \mu_{1}^2 \drangle-\dlangle\mu_{1}\drangle^2
& \dlangle\mu_{1} \mu_{2}\drangle-\dlangle\mu_{1}\drangle\dlangle\mu_{2}\drangle
\\[1ex]
 \dlangle x \mu_{2}\drangle
& \tilde{\Gamma}_{42}
& \dlangle\mu_{1} \mu_{2}\drangle-\dlangle\mu_{1}\drangle\dlangle\mu_{2}\drangle
& \dlangle \mu_{2}^2\drangle -\dlangle\mu_{2}\drangle^2
\end{pmatrix},
\end{align}
where
\bse
\begin{align}
\label{gamma4}
\tilde{\Gamma}_{23} &= \tilde{\Gamma}_{32}
=\dlangle x^2\mu_{1}\drangle+2\chii\dlangle x \mu_{1}\drangle-\frac{1}{12}\dlangle\mu_{1}\drangle,
\\[2ex]
\tilde{\Gamma}_{24} &= \tilde{\Gamma}_{42}
=\dlangle x^2\mu_{2}\drangle+2\chii\dlangle x \mu_{2}\drangle-\frac{1}{12}\dlangle\mu_{2}\drangle.
\end{align}
\ese
As we have indicated above, the elements of the reduced Fisher matrix $\tilde{\Gamma}$
depend on the dimensionless parameter $\chii$ (and on the time-dependent functions
$\mu_1$ and $\mu_2$ introduced in Eqs.\ \eqref{mu1mu2}---they are determined by the motion of the detector with respect to the SSB).
\end{widetext}

\section{Banks of the templates}

To search for the gravitational-wave signal in detector's noise
we need to construct a bank of templates in the space of the parameters $\bxi$
on which the $\F$-statistic [given in Eq.\ \eqref{29}] depends.
The bank of templates is defined by a discrete set of points,
i.e.\ a \emph{grid} in the parameter space chosen in such a way,
that for any possible signal with parameters $\btheta'=(h'_1,h'_2,\bxi')$
there exists a grid point $\bxi$
such that the expectation value of the $\F$-statistic,
$\text{E}_1\{\F[x;\bxi]\}=\text{E}\big\{\F\big[n(t)+h(t;\btheta');\bxi\big]\big\}$,
computed for the signal with parameters $\btheta'$ and for the grid point $\bxi$,
is not less than a certain fixed minimal value,
\emph{assuming that the minimal vaule of the signal-to-noise $\rho$ ratio is \emph{a priori} fixed}.
From Eq.\ \eqref{EF3} we see that this expectation value
depends on the signal-to-noise ratio $\rho$
and on the value $C_0(\bxi,\bxi')$ of the noise autocovariance function
computed for the intrinsic parameters $\bxi$ and $\bxi'$
of the template and the gravitational-wave signal, respectively.

To construct the bank of templates
one thus needs to choose some minimum value $\cmin$ of the autocovariance function $C_0$
and look for such a grid of points that for any signal with parameters $\bxi'$
there exists a grid node $\bxi$ such that the autocovariance $C_0$ computed for the parameters
$\bxi$ and $\bxi'$ is not less than $\cmin$,
\be
\label{bank1}
C_0(\bxi,\bxi') \ge \cmin.
\ee
We employ the linear model of the gravitational-wave signal
for which the autocovariance $C_0(\bxi,\bxi')$ depends on $\bxi$, $\bxi'$
only through the difference $\bxi-\bxi'$,
therefore \eqref{bank1} can be rewritten as
\be
\label{bank2}
C_0(\bxi-\bxi') \ge \cmin.
\ee
\emph{In the rest of this paper we will approximate the autocovariance function $C_0$
by means of the formula \eqref{Ca}},
i.e.\ we will use the approximate equality
\be
\label{bank3}
C_0(\bxi,\bxi') \cong \Ca(\bxi,\bxi').
\ee
By virtue of Eq.\ \eqref{Ca} the inequality \eqref{bank2} can be written as
\be
\label{bank4}
\sum_{k,l=1}^4\tilde{\Gamma}_{kl}\,(\xi_k-\xi'_k)\,(\xi_l-\xi'_l)
\le 1 - \cmin,
\ee
which for the fixed $\bxi$ is fulfilled by all points $\bxi'$ which belong
to an hyperellipsoid with the center located at $\bxi$ and with size
determined by the value of $\cmin$.\footnote{In Appendix \ref{Ametric}
we compare the language used by us to describe the construction of template banks
with the other commonly used language related with the introduction
of a metric in the space of signal's parameters.}

Wa want to find the \emph{optimal} grid fulfilling the requirement \eqref{bank4},
i.e.\ the grid which consists of possibly smallest number of points.
Thus the problem of finding the optimal grid is a kind of \emph{covering} problem,
i.e.\ the problem to cover the $d$-dimensional Euclidean space $\mathbb{R}^d$
(or, in data analysis case, the bounded region of the space)
by the smallest number of \emph{identical} hyperellipsoids.
The thorough exposition of the problem of covering $d$-dimensional Euclidean space
by identical hyperspheres is given in Chap.\ 2 of Ref. \cite{CS99}.

We restrict ourselves to grids which are \emph{lattices},
i.e.\ to grids with nodes which are linear combinations
with integer coefficients of some basis vectors.
If the vectors $(\mathbf{P}_1,\dots,\mathbf{P}_d)$
are the basis vectors of a $d$-dimensional lattice,
then a \emph{fundamental parallelotope} is the subset of the $\mathbb{R}^d$
consisting of the points
\be
\lambda_1 \mathbf{P}_1 + \ldots + \lambda_d \mathbf{P}_d,
\quad
0 \le \lambda_1,\ldots,\lambda_d < 1.
\ee
A fundamental parallelotope is an example of a \emph{fundamental region}
for the lattice, which when repeated many times fills the space with one lattice point in each copy.

The quality of a covering can be expressed by the \emph{covering thickness} $\rho$
which is defined as the average number of hyperellipsoids that contain a point in the space.
For lattice coverings their thickness can be computed as
\be
\rho = \frac{\text{volume of one hyperellipsoid}}{\text{volume of fundamental region}}.
\ee
Thickness of lattice covering of $d$-dimensional space $\mathbb{R}^d$
with identical hyperspheres of radius $R$ reads
\be
\label{ro}
\rho = \frac{2\pi^{d/2}R^{d}}{d\,\Gamma(d/2)|\det\mathsf{E}|},
\ee
where $\mathsf{E}$ is the matrix made of the basis vectors of the lattice.

Another important notion is the \emph{covering radius} of a lattice.
Consider any discrete collection of points
$\mathcal{Q}=\{\mathbf{Q}_1,\mathbf{Q}_2,\ldots\}\subset\mathbb{R}^d$.
The covering radius $R$ of $\mathcal{Q}$ is the least upper bound
for the distance from any point $\mathbf{x}$ of $\mathbb{R}^d$
to the closest point $\mathbf{Q}_i$ of the collection $\mathcal{Q}$,
\be
R := \sup_{\mathbf{x}\in\mathbb{R}^d} \inf_{\mathbf{Q}_i\in\mathcal{Q}} |\mathbf{x}-\mathbf{Q}_i|.
\ee
Then identical hyperspheres of radius $R$ centered at the points of $\mathcal{Q}$ will cover $\mathbb{R}^d$,
and no hyperspheres of radius smaller than $R$ will cover it.
Around each point $\mathbf{Q}_i$ one defines its \emph{Voronoi cell}, $V(\mathbf{Q}_i)$,
which consists of those point of $\mathbb{R}^d$
that are at least as close to $\mathbf{Q}_i$ as to any other $\mathbf{Q}_j$,
\be
V(\mathbf{Q}_i) := \{\mathbf{x}\in\mathbb{R}^d:
|\mathbf{x}-\mathbf{Q}_i|\le|\mathbf{x}-\mathbf{Q}_j|\,\text{for all $j$}\}.
\ee
The Voronoi cell is sometimes called a Wigner-Seitz cell
(and also the span of a template
in the literature on analysis of gravitational-wave signals \cite{1999-Owen-Sathyaprakash}).
The interiors of the Voronoi cells are disjoint. Each face of the Voronoi cell
lies in the hyperplane midway between two neighboring points $\mathbf{Q}_i$. Voronoi cells
are convex polytopes whose union is the whole $\mathbb{R}^d$.
If the collection $\mathcal{Q}$ forms a lattice, then all the Voronoi cells are congruent.

\subsection{Constraints}

As we have already mentioned in Sec.\ I
we are interested in searches for gravitational-wave signals
with very large number of grid points in the parameter space.
Then the time needed to compute the $\F$-statistic for all grid nodes is long
and we want to speed up this computation by employing the FFT algorithm.
As the FFT algorithm computes the values of the DFT of a time series
for a certain set of discrete frequencies (the Fourier frequencies),
it will be possible to use the FFT algorithm
if the frequency coordinates of grid points
will all coincide with the Fourier frequencies.
We want to construct grids which fulfill this requirement.

Le the data collected by a detector form a sequence of $N$ samples
\be
x_u, \quad u=1,\ldots,N,
\ee
and let the sampling-in-time period be $\Delta{t}$. Then the DFT algorithm calculates
the Fourier transform of the data with the frequency resolution $\Delta{f}=1/(N\Delta{t})$.
The resolution of the dimensionless frequency parameter $\omega_0$
[introduced in Eq.\ \eqref{omegak}] is thus
\be
\label{Domega0}
\Delta{\omega_0} = 2\pi\To \Delta{f} = 2\pi,
\ee
because $N\Delta{t}=\To$.

It is possible to modify the DFT algorithm in such a way,
that the frequency resolution \eqref{Domega0} changes.
There exisit two types of such modifications
(see Appendix in Ref.\ \cite{PJP2011} for more details):
(i) zero-padding of the data, which makes the DFT more dense;
(ii) folding of the data, which diminishes the frequency resolution.
If we add $N_{\text{FFT}}-N$ zeros to the $N$ samples,
so we make the Fourier transform of $N_{\text{FFT}}$ data points,
then the frequency resolution is
\be
\label{FFTwiaz1}
\Delta\omega_{0} = 2\pi\,\frac{N}{N_{\text{FFT}}}.
\ee
The case $N_{\text{FFT}}=2N$ corresponds to usual zero-padding of the data,
where for $N$ data points we add $N$ zeros.
If we in turn fold $N$-point data stream $p$ times
(so we finally have $N/2^p$ data points),
then the frequency resolution changes to
\be
\label{FFTwiaz2}
\Delta{\omega_0} = 2^p\times2\pi, \quad p=1,2,\ldots.
\ee

Let $(\mathbf{P}_1,\mathbf{P}_2,\mathbf{P}_3,\mathbf{P}_4)$
be the basis vectors of a 4-dimensional lattice we consider.
To use the FFT algorithm we need such grid that all nodes
can be arranged along straight lines parallel to the $\omega_{0}$ axis
and the distance between neighboring nodes along these lines
must be equal to the frequency resolution of the FFT algorithm.
We thus require that (say) the first basis vector of the grid
we are looking for is of the form\footnote{
In the rest of the paper we treat all 4-vectors as \emph{column} $4\times1$ matrices.
We will also use matrix notation with
superscript ``$\mathsf{T}$'' denoting matrix transposition
and ``$\cdot$'' denoting matrix multiplication.}
\be
\label{wiazP1}
\mathbf{P}_1 = (\Delta\omega_0,0,0,0)^\mathsf{T},
\ee
where $\Delta\omega_0$ is the frequency resolution of the FFT algorithm
we want to use [it is given in Eqs.\ \eqref{FFTwiaz1} or \eqref{FFTwiaz2}].

There is another important constraint to be met
by grids employed in all-sky searches.
It is related to the reduction of the computational time needed to resample the data
to the so called \emph{barycentric time}
(see e.g.\ Sec.\ III D in \cite{JKS98}, Sec.\ V A in \cite{ABJPK10}, \cite{PSDB2010},
and Sec.\ 6.2 in \cite{2014--LSC-VC--CQG-b}).
Numerically accurate resampling is computationally demanding,
therefore it is necessary to construct such grids,
that the resampling is needed only once per sky position for all spindown values.
To meet this constraint we require that the 3rd and the 4th component
of the (say) second basis vector vanish:
\be
\label{wiazP2}
\mathbf{P}_2 = (P_{21},P_{22},0,0)^\mathsf{T}.
\ee

\subsection{Replacing hyperellipsoid-coverings by hypersphere-coverings}

It is convenient to replace the problem of finding the optimal covering of space by identical hyperellipsoids
by the problem of finding the optimal covering of space by identical hyperspheres.
Let us denote the original space of grid parameters by $\Omega$
($\Omega\subset\mathbb{R}^4$, $\btau\in\Omega$)
and let the space of the transformed grid parameters be $\Omega'$
($\Omega'\subset\mathbb{R}^4$, $\btau'\in\Omega'$).
We are looking for the linear transformation
\be
\label{501}
\btau' = \mathsf{F}(\chii,\cmin) \cdot \btau,
\ee
which transforms the hyperellipsoid of the constant value of the autocovariance function
into the hypersphere of unit radius.
This hyperellipsoid of the constant value equal to $\cmin$
is determined by the equation [see Eq.\ \eqref{bank4}]
\be
\label{502}
\btau^\top \cdot \tilde{\Gamma}(\chii)\cdot \btau - (1-\cmin) = 0,
\ee
whereas the equation of the unit hypersphere in the $\Omega'$ space reads
\be
\label{503}
{\btau'}^\mathsf{T}\cdot\btau' - 1 = 0.
\ee
After substituting \eqref{501} into \eqref{502} we get
\be
\label{504}
\btau^\mathsf{T} \cdot \mathsf{F}(\chii,\cmin)^\mathsf{T} \cdot \mathsf{F}(\chii,\cmin)\cdot \btau-1=0.
\ee
Equations \eqref{504} and \eqref{502} describe the same hyperellipsoid if and only if 
\be
\label{505}
\mathsf{F}(\chii,\cmin)^\mathsf{T} \cdot \mathsf{F}(\chii,\cmin)
= \frac{1}{R(\cmin)^{2}}\tilde{\Gamma}(\chii),
\ee
where
\be
\label{506}
\quad R(\cmin) \coloneqq \sqrt{1-\cmin}
\ee
is the average radius of the hyperellipsoid \eqref{502}.

The Fisher matrix $\tilde{\Gamma}$ is symmetric and
[what can be shown by means of Eq.\ \eqref{F5}]
it is strictly positive definite,
i.e.\ $\btau^\top\cdot\tilde{\Gamma}\cdot\btau>0$
for any $\btau\ne\mathbf{0}$.
Therefore the equation \eqref{505} can be interpreted
as the \emph{Cholesky decomposition} of the matrix $\tilde{\Gamma}/R^2$,
which states that there  exists the unique \emph{upper triangular matrix} $F$
(with strictly positive diagonal elements) fulfilling Eq.\ \eqref{505}.
In the rest of the paper we will assume that the matrix $F$
is the result of the Cholesky decomposition (so it is an upper triangular matrix).

The elements of the upper triangular matrix $\mathsf{F}$ depend on the parameters $\chii$ and $\cmin$
and, through the evaluation of time averages needed to compute
the elements of the Fisher matrix $\tilde{\Gamma}$ [see Eq.\ \eqref{F5}],
on the position vector of the detector with respect to the SSB during observational interval.
Making use of Eq.\ \eqref{F5} it is easy to show that the $(1,1)$ element of the matrix $\mathsf{F}$
depends only on $\cmin$ and it is equal to
\be
\label{F11}
F_{11}(\cmin) = \frac{1}{2\sqrt{3(1-\cmin)}}.
\ee

When the basis vectors of the grid in $\Omega'$ space are found,
one transform them into $\Omega$ space by means of the inverse matrix $\mathsf{F}^{-1}$.
If $\mathsf{C}'$ is the generating matrix of the grid in $\Omega'$ space,
then the generating matrix $\mathsf{C}$ of the corresponding grid in $\Omega$ space can be computed as
\be
\label{odo}
\mathsf{C} = \mathsf{C}' \cdot (\mathsf{F}^{-1}(\chii,\cmin))^\mathsf{T}.
\ee
Both matrices $\mathsf{C}$ and $\mathsf{C}'$ are lower triangular
(the matrix $\mathsf{F}^{-1}$ is upper triangular so its transpose is lower triangular as well).

\section{Construction of grids}

We will construct in the present section two families of grids
which meet the two constraints \eqref{wiazP1} and \eqref{wiazP2}.
As far as we know, the general solution to the covering problem with constraints is not known.
Our grids constructed below are generally better than the grids previously proposed,
but this of course does not exclude possibility that better grids still might exist.

We start from transforming the first basis vector $\mathbf{P}_1$
of the grid we are looking for,
which is fixed and given in Eq.\ \eqref{wiazP1},
from $\Omega$ into $\Omega'$ space,
\be
\mathbf{P}'_1 \coloneqq
F(\chii,\cmin)\cdot(\Delta\omega_0,0,0,0)^\mathsf{T}.
\ee
Because the matrix $F$ is upper triangular,
this leads to
\be
\label{wiaz1}
\mathbf{P}'_1 = (\Delta\omega_{0}',0,0,0)^\mathsf{T},
\ee
where, by means of Eq.\ \eqref{F11},
the length $\Delta\omega_{0}'$ of the basis vector $\mathbf{P}'_1$
in $\Omega'$ space equals
\be
\label{omegavsomega}
\Delta\omega_0' = \frac{\Delta\omega_0}{2\sqrt{3(1-\cmin)}}.
\ee
We also require that the second basis vector $\mathbf{P}_2$
of the grid in $\Omega$ space
fulfills the constraint \eqref{wiazP2}.
Therefore in $\Omega'$ space the second basis vector
has to have its 3rd and 4th component equal to zero:
\be
\label{wiaz2}
\mathbf{P}'_2 \coloneqq
F(\chii,\cmin)\cdot(P_{21},P_{22},0,0)^\mathsf{T}
= (P'_{21},P'_{22},0,0)^\mathsf{T}.
\ee
All grids constructed in the present section in $\Omega'$ space
fulfill the constraints \eqref{wiaz1}--\eqref{omegavsomega} and \eqref{wiaz2}.
Constructions of grids depend [through Eq.\ \eqref{omegavsomega}]
only on the values of $\Delta\omega_{0}$ and $\cmin$. 
All grids constructed in $\Omega'$ space
are built up from \emph{hyperspheres of unit radius}.

We will give below explicit numerical results for grids
computed for some exemplary values of the parameters of the search.
As the number of data points we take
\be
\label{no.of.data.points}
N=344656.
\ee
This number corresponds to the observational interval of 2 sidereal days
sampled with the time period of 0.5~s.
We consider the following lengths of the FFT:
\be
\label{nfft}
N_\text{FFT} = 2^{19},\ 2^{20},\ 2^{21}.
\ee
For these values of the search parameters
the frequency resolution of the FFT according to Eq.\ \eqref{FFTwiaz1} reads,
respectively,
\be
\label{domega0}
\Delta\omega_0 \cong
4.13044,\
2.06522,\
1.03261.
\ee
We will consider the minimum value of the autocovariance $\cmin$
taken from the interval $\left<0.70,\,0.999\right>$,
and sometimes as a reference value of $\cmin$ we take $\cmin=0.75$.
For this reference value of $\cmin$ the length $\Delta\omega_{0}'$
of the basis vector $\mathbf{P}'_1$
computed by means of Eq.\ \eqref{omegavsomega}
for the frequency resolution given in Eq.\ \eqref{nfft}
equals, respectively,
\be
\label{wiaz2n}
\Delta\omega_{0}' \cong
2.38471,\
1.19235,\
0.596177.
\ee

\subsection{Optimal $A_4^\star$ lattice}

Our constructions are based on some deformations
of the 4-dimensional optimal lattice covering of space $\mathbb{R}^4$.
This is so called $A_{4}^{\star}$ lattice \cite{CS99},
which is generated by the following matrix
(made of the basis vectors $\mathbf{v}_1,\ldots,\mathbf{v}_4$
with components arranged into rows of the matrix):
\be
\label{v1}
\mathsf{V} = \left(\begin{array}{c}
\mathbf{v}_{1}^\mathsf{T}\\[1ex]
\mathbf{v}_{2}^\mathsf{T}\\[1ex]
\mathbf{v}_{3}^\mathsf{T}\\[1ex]
\mathbf{v}_{4}^\mathsf{T}
\end{array}\right)
= \left(
\begin{array}{cccc}
\sqrt{5} & 0 & 0 & 0 \\[0.5ex]
\frac{\sqrt{5}}{2} & \frac{\sqrt{15}}{2} & 0 & 0 \\[0.5ex]
\frac{\sqrt{5}}{2} & \frac{\sqrt{\frac{5}{3}}}{2} & \sqrt{\frac{10}{3}} & 0 \\[0.5ex]
-\frac{\sqrt{5}}{2} & -\frac{\sqrt{\frac{5}{3}}}{2} & -\frac{\sqrt{\frac{5}{6}}}{2} & -\frac{1}{2 \sqrt{2}}
\end{array} \right).
\ee
This matrix generates covering of space by hyperspheres of unit radii.
Euclidean lengths of the basis vectors are:
\be
\label{v2}
|\mathbf{v}_{1}|=|\mathbf{v}_{2}|=|\mathbf{v}_{3}|=\sqrt{5},
\quad
|\mathbf{v}_{4}|=\sqrt{2}.
\ee
Thickness of the optimal lattice covering $A_{4}^{\star}$ is
\be
\label{ro_A4*}
\rho_{A_{4}^{\star}}
= \frac{\pi^{2}}{2\,|\det \mathsf{V}|}
= \frac{2\pi^2}{5\sqrt{5}}
\cong 1.765529.
\ee

In the construction of the constrained grids given below
the crucial role is played by the lengths of the lattice vectors
for the optimal lattice $A_{4}^{\star}$,
i.e.\ the lengths of the vectors joining any two nodes of the lattice.
The first 119 smallest \emph{squares} of the lengths, in ascending order,
are listed in Appendix \ref{ALengths}.

\subsection{Grids $S_1$ for $\cmin<\cmin^*$}
\label{subsecS1a}

Our first construction will be valid
(for the reasons explained at the end of the present subsection),
for the given value of the frequency resolution $\Delta\omega_0$,
only in the case when the minimum value $\cmin$ of the autocovariance is less than $\cmin^*$,
where $\cmin^*$ is defined in Eq.\ \eqref{cmin*} below.

We begin the construction from replacing
the basis vectors $\mathbf{v}_a\,(a=1,\ldots,4)$ of the lattice $A_{4}^{\star}$
by the new basis consisting of the following vectors:
\be
\label{m1}
\begin{array}{l}
\mathbf{m}_{1} \coloneqq \mathbf{v}_{1}+\mathbf{v}_{4},\\[1ex]
\mathbf{m}_{2} \coloneqq \mathbf{v}_{2}+\mathbf{v}_{4},\\[1ex]
\mathbf{m}_{3} \coloneqq \mathbf{v}_{3}+\mathbf{v}_{4},\\[1ex]
\mathbf{m}_{4} \coloneqq \mathbf{v}_{4}.
\end{array}
\ee
Generating matrix of the lattice $A_{4}^{\star}$
with rows made of these vectors reads
\be
\label{m2}
\left(\begin{array}{c}
\mathbf{m}_{1}^\mathsf{T}\\[1ex]
\mathbf{m}_{2}^\mathsf{T}\\[1ex]
\mathbf{m}_{3}^\mathsf{T}\\[1ex]
\mathbf{m}_{4}^\mathsf{T}
\end{array}\right)
= \left(\begin{array}{cccc}
 \frac{\sqrt{5}}{2} & -\frac{\sqrt{\frac{5}{3}}}{2} & -\frac{\sqrt{\frac{5}{6}}}{2} & -\frac{1}{2 \sqrt{2}} \\
 0 & \sqrt{\frac{5}{3}} & -\frac{\sqrt{\frac{5}{6}}}{2} & -\frac{1}{2 \sqrt{2}} \\
 0 & 0 & \frac{\sqrt{\frac{15}{2}}}{2} & -\frac{1}{2 \sqrt{2}} \\
 -\frac{\sqrt{5}}{2} & -\frac{\sqrt{\frac{5}{3}}}{2} & -\frac{\sqrt{\frac{5}{6}}}{2} & -\frac{1}{2 \sqrt{2}}
\end{array}\right).
\ee
Lengths of all vectors $\mathbf{m}_a$ are the same,
\be
\label{m3}
|\mathbf{m}_{1}|=|\mathbf{m}_{2}|=|\mathbf{m}_{3}|=|\mathbf{m}_{4}|=\sqrt{2}.
\ee
Cosine of the angle and the angle itself between any two of the vectors $\mathbf{m}_a$ equal
\be
\label{m4}
\cos\beta = -1/4,
\quad
\beta \cong 1.82348.
\ee

Let us make the tails of the vectors $\mathbf{m}_a$ to coincide with the origin of the coordinate system.
Equations \eqref{m3} and \eqref{m4} imply then that the vectors $\mathbf{m}_a$ coincide with
the side edges of a 4-dimensional simplex whose base is a 3-dimensional regular tetrahedron
(with vertexes determined by the heads of the vectors $\mathbf{m}_a$).
The vector
\be
\label{m5}
\mathbf{h}_{1} \coloneqq 
\frac{1}{4}(\mathbf{m}_{1}+\mathbf{m}_{2}+\mathbf{m}_{3}+\mathbf{m}_{4})
= \left(0,0,0,-\frac{1}{2 \sqrt{2}}\right)^\mathsf{T}.
\ee
coincides with the height of this simplex.

Grid points of the $A_4^\star$ lattice are determined by the vectors
\be
\label{m6}
\mathbf{w}(i,j,k,l) \coloneqq i\,\mathbf{m}_{1}+j\,\mathbf{m}_{2}+k\,\mathbf{m}_{3}+l\,\mathbf{m}_{4},
\ee
where $i,j,k,l$ are any integers. We look for integers $i,j,k,l$ such that the vector $\mathbf{w}(i,j,k,l)$
has length not less than $\Delta\omega_0'$ and as close to $\Delta\omega_0'$ as possible.
There are many different choices leading to vectors of the same lengths. In this subsection
we choose $i=j=k=l=1$ what gives the following vector (parallel to the vector $\mathbf{h}_1$):
\be
\label{m7}
\mathbf{m}_{5}=\big(0,0,0,-\sqrt{2}\big)^\mathsf{T}.
\ee
After replacing in the matrix \eqref{m2} the vector $\mathbf{m}_{4}$ by $\mathbf{m}_{5}$,
we get another matrix generating the lattice $A_{4}^{\star}$,
\be
\label{m8}
\left(
\begin{array}{cccc}
 \frac{\sqrt{5}}{2} & -\frac{\sqrt{\frac{5}{3}}}{2} & -\frac{\sqrt{\frac{5}{6}}}{2} & -\frac{1}{2 \sqrt{2}} \\
 0 & \sqrt{\frac{5}{3}} & -\frac{\sqrt{\frac{5}{6}}}{2} & -\frac{1}{2 \sqrt{2}} \\
 0 & 0 & \frac{\sqrt{\frac{15}{2}}}{2} & -\frac{1}{2 \sqrt{2}} \\
 0 & 0 & 0 & -\sqrt{2}
\end{array}
\right).
\ee
Now in three steps we will transform the matrix \eqref{m8}
into a lower triangular matrix generating the same $A_{4}^{\star}$ lattice.
The height of the simplex built up from the basis vectors related to this new matrix
will be parallel to the $\omega_0'$ axis.

\begin{enumerate}
\item
We exchange the order of the basis vectors: $1\leftrightarrow4$, $2\leftrightarrow3$.
This leads to the matrix
\be
\label{516_1}
\left(\begin{array}{cccc}
 0 & 0 & 0 & -\sqrt{2} \\
 0 & 0 & \frac{\sqrt{\frac{15}{2}}}{2} & -\frac{1}{2 \sqrt{2}} \\
 0 & \sqrt{\frac{5}{3}} & -\frac{\sqrt{\frac{5}{6}}}{2} & -\frac{1}{2 \sqrt{2}} \\
 \frac{\sqrt{5}}{2} & -\frac{\sqrt{\frac{5}{3}}}{2} & -\frac{\sqrt{\frac{5}{6}}}{2} & -\frac{1}{2 \sqrt{2}}
\end{array}\right).
\ee
\item
We transform the basis vectors by means of the reflection with respect to the hyperplanes
$\omega_{0}'-\alpha_{2}'=0$ [with normal vector $(-1,0,0,1)^\mathsf{T}$]
and $\omega_{1}'-\alpha_{1}'=0$ [with normal vector $(0,-1,1,0)^\mathsf{T}$].
This leads to the exchange of the columns: $1\leftrightarrow 4$, $2\leftrightarrow 3$,
and gives the matrix
\be
\label{516_2}
\left(\begin{array}{cccc}
 -\sqrt{2} & 0 & 0 & 0 \\
 -\frac{1}{2 \sqrt{2}} & \frac{\sqrt{\frac{15}{2}}}{2} & 0 & 0 \\
 -\frac{1}{2 \sqrt{2}} & -\frac{\sqrt{\frac{5}{6}}}{2} & \sqrt{\frac{5}{3}} & 0 \\
 -\frac{1}{2 \sqrt{2}} & -\frac{\sqrt{\frac{5}{6}}}{2} & -\frac{\sqrt{\frac{5}{3}}}{2} & \frac{\sqrt{5}}{2}
\end{array}\right).
\ee
\item
It is convenient to have generating matrix with all diagonal elements positive.
To achieve this we reflect the basis vectors of the matrix \eqref{516_2}
with respect to the hyperplane $\omega_{0}'=0$.
We get the generating matrix $\mathsf{N}$ built up from the basis vectors
$\mathbf{n}_a$ ($a=1,2,3,4$):
\be
\label{N}
\mathsf{N} = \left(\begin{array}{c}
\mathbf{n}_{1}^\mathsf{T}\\[1ex]
\mathbf{n}_{2}^\mathsf{T}\\[1ex]
\mathbf{n}_{3}^\mathsf{T}\\[1ex]
\mathbf{n}_{4}^\mathsf{T}
\end{array}\right)
= \left(\begin{array}{cccc}
 \sqrt{2} & 0 & 0 & 0 \\
 \frac{1}{2 \sqrt{2}} & \frac{\sqrt{\frac{15}{2}}}{2} & 0 & 0 \\
 \frac{1}{2 \sqrt{2}} & -\frac{\sqrt{\frac{5}{6}}}{2} & \sqrt{\frac{5}{3}} & 0 \\
 \frac{1}{2 \sqrt{2}} & -\frac{\sqrt{\frac{5}{6}}}{2} & -\frac{\sqrt{\frac{5}{3}}}{2} & \frac{\sqrt{5}}{2}
\end{array}\right).
\ee
\end{enumerate}

Finally we \emph{deform} the optimal lattice $A_{4}^{\star}$ in the following way:
we squeeze this lattice in the direction of the $\omega_{0}'$ axis
by the factor $q\le1$, where
\be
\label{q}
q \coloneqq \Delta\omega_{0}'/\sqrt{2}.
\ee
The resulting grid is generated by the matrix
\be
\label{C1}
\mathsf{C}_1 = \mathsf{N} \cdot \mathsf{Q},
\ee
where $\mathsf{Q}$ is the diagonal matrix with elements
\be
\label{Q}
\mathsf{Q} \coloneqq \left(\begin{array}{cccc}
 q & 0 & 0 & 0 \\
 0 & 1 & 0 & 0 \\
 0 & 0 & 1 & 0 \\
 0 & 0 & 0 & 1
\end{array}\right).
\ee
The elements of the generating matrix $\mathsf{C}_{1}$ thus read
\be
\label{C1sy}
\mathsf{C}_1 = \left(\begin{array}{cccc}
 \Delta\omega_{0}' & 0 & 0 & 0 \\[1ex]
 \frac{1}{4}\Delta\omega_{0}' & \frac{\sqrt{\frac{15}{2}}}{2} & 0 & 0 \\[1ex]
 \frac{1}{4}\Delta\omega_{0}' & -\frac{\sqrt{\frac{5}{6}}}{2} & \sqrt{\frac{5}{3}} & 0 \\[1ex]
 \frac{1}{4}\Delta\omega_{0}' & -\frac{\sqrt{\frac{5}{6}}}{2} & -\frac{\sqrt{\frac{5}{3}}}{2} & \frac{\sqrt{5}}{2}
\end{array}\right).
\ee
Obviously the first two rows of this matrix form the vectors
which meet the two constraints \eqref{wiaz1} and \eqref{wiaz2}.
Let us denote by $S_1$ the grid generated by the matrix $\mathsf{C}_1$.

The thickness of the grid $S_1$ expressed in terms of $\Delta\omega_{0}'$ is equal
\be
\rho_{S_{1}} = \frac{\pi^{2}}{2\,|\det\mathsf{C}_{1}|}
= \left(\frac{2}{5}\right)^{3/2} \frac{\pi^2}{\Delta\omega_{0}'}.
\ee
By means of Eq. \eqref{omegavsomega} $\Delta\omega_{0}'$ can be expressed
by search parameters $\Delta\omega_{0}$ and $\cmin$.
Then the thickness of the grid $S_1$ can be written as
\be
\rho_{S_1} = \frac{4}{5}\sqrt{\frac{6}{5}} \pi^2
\frac{\sqrt{1-\cmin}}{\Delta\omega_{0}}.
\ee

As an example let us take $\cmin=0.75$ and $\Delta\omega_0\cong2.06522$.
Then $\Delta\omega_0'\cong1.19235$,
the coefficient $q$ of Eq.\ \eqref{q} equals $q\cong0.843122$,
and the matrix $\mathsf{C}_1$ has elements
\begin{widetext}
\be
\label{S1a}
\mathsf{C}_{1} = \left(\begin{array}{cccc}
 1.192353885 & 0 & 0 & 0 \\
 0.2980884713 & 1.369306394 & 0 & 0 \\
 0.2980884713 & -0.4564354646 & 1.290994449 & 0 \\
 0.2980884713 & -0.4564354646 & -0.6454972244 & 1.118033989
\end{array}\right),
\ee
\end{widetext}
whereas the thickness of the grid $S_{1}$ is
\be
\label{ro_s1}
\rho_{S_{1}} \cong 2.094038.
\ee

\begin{figure}
\begin{center}
\includegraphics[scale=0.6]{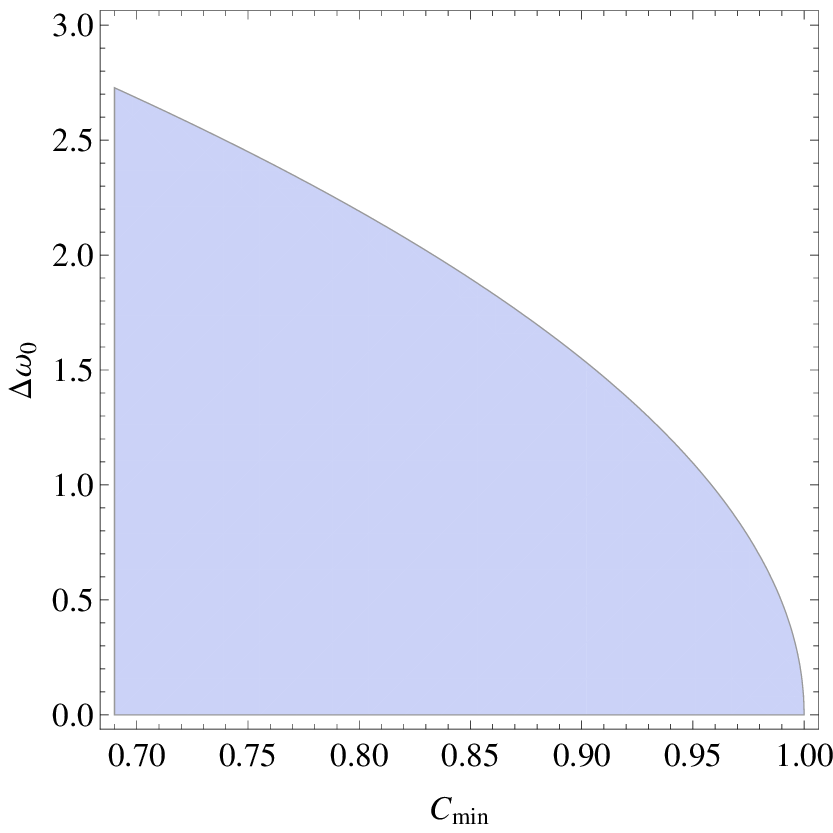}
\caption{\label{S1range}
The shaded region in the plane $(\cmin,\Delta\omega_0)$ is made of the points
for which the construction of the grid $S_1$ described in Sec.\ \ref{subsecS1a} is possible.
Outside this region the construction from Sec.\ \ref{subsecS1b} can be used.}
\end{center}
\end{figure}

Equations \eqref{q} and \eqref{omegavsomega} imply
that the construction described above works only
when $\Delta\omega_{0}'\le\sqrt{2}$, i.e.\
when the resolution $\Delta\omega_{0}$ of the parameter
$\omega_{0}$ and $\cmin$ fulfill the inequality
(see Fig.\ \ref{S1range})
\be
\Delta\omega_0 \le 2\sqrt{6}\sqrt{1-\cmin}.
\ee
Moreover, there is no need to deform the optimal lattice $A_{4}^{\star}$
in the case when $\Delta\omega_{0}'=\sqrt{2}$, i.e.\ when the above
inequality becomes the equality. For the fixed value of the parameter $\Delta\omega_0$
let us denote by $\cmin^*$ the limiting value of the parameter $\cmin$.
Then
\be
\label{cmin*}
\Delta\omega_0 = 2\sqrt{6}\sqrt{1-\cmin^*}
\quad\text{or}\quad
\cmin^* = 1 - \frac{1}{24}\Delta\omega_0^2.
\ee
When $\Delta\omega_0\cong 4.13044,\,2.06522,\,1.03261$,
this condition holds for $\cmin^*\cong0.289146,\,0.822287,\,0.955572$,
respectively.
If for the given value of the resolution $\Delta\omega_{0}$
one wants to consider $\cmin>\cmin^*$,
then the contruction of the grid described in the next subsection can be used.

\subsection{Grids $S_1$ for $\cmin>\cmin^*$}
\label{subsecS1b}

In this subsection we will describe construction of grids $S_1$
valid also when $\cmin>\cmin^*$.
We start again from looking for such basis vector of the optimal lattice $A_4^\star$
which has length as close to $\Delta\omega_{0}'$ as possible
(but not less than $\Delta\omega_{0}'$).
Making use of the basis vectors $\mathbf{n}_a$ of the lattice $A_4^\star$
[they are rows of the matrix $\mathsf{N}$ from Eq.\ \eqref{N}],
grid points of the lattice can be written in the form
\be
\label{wektor2}
\mathbf{q}(i,j,k,l) \coloneqq i\,\mathbf{n}_{1}+j\,\mathbf{n}_{2}+k\,\mathbf{n}_{3}+l\,\mathbf{n}_{4},
\ee
where $i,j,k,l$ are integers.
We require that the chosen vector is such that one can take it
and some three out of four basis vectors $\mathbf{n}_a$
to form a new basis of the lattice $A_4^\star$.
Let us denote the chosen vector
(with the length closest to $\Delta\omega_0'$)
by $\mathbf{q}$ and let its components be
\be
\label{wektorq}
\mathbf{q} = (a,b,c,d)^\mathsf{T}.
\ee
We will rotate this vector to make it parallel to the $\Delta\omega_0'$ axis.

Let us first find any angle $\beta$ satisfying equations
\be
\cos\beta = \frac{c}{\sqrt{b^2+c^2}},\quad
\sin\beta = \frac{b}{\sqrt{b^2+c^2}},
\ee
and let us introduce the matrix $\mathsf{R}_1(\beta)$
describing rotation in the 2-plane $(\omega_1',\alpha_1')$ by the angle $\beta$,
\be
\label{ob}
\mathsf{R}_1(\beta) \coloneqq \left(\begin{array}{cccc}
 1 & 0 & 0 & 0 \\
 0 & \cos\beta & -\sin\beta & 0 \\
 0 & \sin\beta & \cos\beta & 0 \\
 0 & 0 & 0 & 1
\end{array}\right).
\ee
Then we define
\be
\mathbf{q}_1 \coloneqq \mathsf{R}_1(\beta)\cdot\mathbf{q}
= (a,0,\sqrt{b^2+c^2},d)^\mathsf{T}.
\ee
Next we rotate the vector $\mathbf{q}_1$
in the 2-plane $(\alpha_1',\alpha_2')$ by the angle $\gamma$ such that
\be
\cos\gamma = \frac{d}{\sqrt{b^2+c^2+d^2}},\quad
\sin\gamma = \frac{\sqrt{b^2+c^2}}{\sqrt{b^2+c^2+d^2}}.
\ee
To do this we introduce the rotation matrix
\be
\label{og}
\mathsf{R}_2(\gamma) = \left(\begin{array}{cccc}
 1 & 0 & 0 & 0 \\
 0 & 1 & 0 & 0 \\
 0 & 0 & \cos\gamma & -\sin\gamma \\
 0 & 0 & \sin\gamma & \cos\gamma 
\end{array}\right)
\ee
and define the vector
\be
\mathbf{q}_2 \coloneqq \mathsf{R}_2(\gamma)\cdot\mathbf{q}_1
= (a,0,0,\sqrt{b^2+c^2+d^2})^\mathsf{T}.
\ee
Finally we rotate the vector $\mathbf{q}_2$
in the 2-plane $(\omega_0',\alpha_2')$
employing the matrix
\be
\label{od}
\mathsf{R}_3(\delta) = \left(\begin{array}{cccc}
 \cos\delta & 0 & 0 & -\sin\delta \\
 0 & 1 & 0 & 0 \\
 0 & 0 & 1 & 0 \\
 \sin\delta & 0 &  & \cos\delta
\end{array} \right),
\ee
where the angle $\delta$ is such that
(here $|\mathbf{q}|=\sqrt{a^2+b^2+c^2+d^2}$)
\be
\cos\delta = \frac{a}{|\mathbf{q}|},\quad
\sin\delta = -\frac{\sqrt{b^2+c^2+d^2}}{|\mathbf{q}|}.
\ee
We define
\be
\label{wektorq5}
\mathbf{q}_3 \coloneqq \mathsf{R}_3(\delta)\cdot\mathbf{q}_2
= (|\mathbf{q}|,0,0,0)^\mathsf{T}.
\ee
The vector $\mathbf{q}_3$ is obviously parallel to the $\omega_0'$ axis.

We now replace one of the basis vectors $\mathbf{n}_a$
of the lattice $A_4^\star$ by the vector $\mathbf{q}$. 
The vector $\mathbf{q}$ is a linear combination $\sum_a{i_a\mathbf{n}_a}$
such that at least for one label $b$ we have $i_b=1$.
We replace $\mathbf{n}_b$ by the vector $\mathbf{q}$
and form a new basis for the lattice $A_4^\star$
made of $\mathbf{q}$ and the rest of the vectors $\mathbf{n}_a$.
Then we apply the triple rotation
$\mathsf{R}_1(\beta)\cdot\mathsf{R}_2(\gamma)\cdot\mathsf{R}_3(\delta)$
to this new basis. Arranging components of the vectors
into rows we get the generating matrix of the lattice $A_4^\star$ of the form
\be
\label{N0}
\mathsf{N}_{0}=
\left(\begin{array}{cccc}
 |\mathbf{q}| & 0 & 0 & 0 \\
 p_{1} & p_{2} & p_{3} & p_{4} \\
 r_{1} & r_{2} & r_{3} & r_{4} \\
 s_{1} & s_{2} & s_{3} & s_{4}
\end{array} \right).
\ee
Making two further rotations in the 2-planes $(\omega_1',\alpha_1')$ and $(\alpha_1',\alpha_2')$
we can make the $p_{3}$ and $p_{4}$ components of the second basis vector to be zero.
After this the generating matrix of the lattice $A_4^\star$ has the form
\be
\label{N1}
\mathsf{N}_{1} = \left(\begin{array}{cccc}
 |\mathbf{q}| & 0 & 0 & 0 \\
 p'_{1} & p'_{2} & 0 & 0 \\
 r'_{1} & r'_{2} & r'_{3} & r'_{4} \\
 s'_{1} & s'_{2} & s'_{3} & s'_{4}
\end{array}\right).
\ee

Now we squeeze the optimal lattice $A_{4}^{\star}$
in the direction of the $\omega_{0}'$ axis.
To do this we employ the diagonal matrix
\be
\label{Q1}
\mathsf{Q}_{1} = \left(\begin{array}{cccc}
 q_1 & 0 & 0 & 0 \\
 0 & 1 & 0 & 0 \\
 0 & 0 & 1 & 0 \\
 0 & 0 & 0 & 1
\end{array}\right),
\ee
where the squeezing factor
\be
\label{q2}
q_1 \coloneqq \Delta\omega_{0}'/|\mathbf{q}|
\ee
is not greater than 1.
The generating matrix of the grid $S_{1}$ reads
\be
\label{C1b}
\mathsf{C}_{1} = \mathsf{N}_{1} \cdot \mathsf{Q}_{1}.
\ee
The first basis vector of this grid
(which components form the first row of the matrix $\mathsf{C}_1$)
coincides with the constraint \eqref{wiaz1}
and the second one (with components taken from the second row of the matrix $\mathsf{C}_1$)
fulfills the constraint \eqref{wiaz2}.

For the chosen vector $\mathbf{q}$ the construction described above
leads to covering only for $\Delta\omega_{0}'\le|\mathbf{q}|$.
In the limiting case when $\Delta\omega_{0}'=|\mathbf{q}|$
the grid $S_1$ coincides with the optimal grid $A_{4}^{\star}$.
The thickness of the grid $S_1$ equals
\be
\label{rhoS1b1}
\rho_{S_{1}} = \frac{\pi^{2}}{2\,|\det\mathsf{C}_{1}|}
= \frac{2\pi^2}{5\sqrt{5}} \frac{|\mathbf{q}|}{\Delta\omega_{0}'},
\ee
or, if one expresses $\Delta\omega_{0}'$ by search parameters $\Delta\omega_0$ and $\cmin$,
\be
\label{rhoS1b2}
\rho_{S_{1}} = \frac{4\sqrt{3}\pi^2}{5\sqrt{5}}
\frac{|\mathbf{q}|\sqrt{1-\cmin}}{\Delta\omega_0}.
\ee

\subsection{Grids $S_2$}
\label{S2grids}

In this subsection we introduce a new family $S_2$ of grids,
which in general have smaller than $S_1$ thicknesses
for smaller values of the parameter $\Delta\omega_0'$ [see Eq.\ \eqref{omegavsomega}],
what can easily be seen in Fig.\ \ref{s1s2omega1}.
Grids $S_2$ will thus have in general advantage over grids $S_1$
for smaller values of the parameter $\cmin$ and/or for finer frequency spacings $\Delta\omega_0$.

The family of grids $S_1$ was obtained in the two previous subsections
as the result of squeezing the optimal 4-dimensional lattice $A_4^\star$.
Our starting point in the constructions of grids $S_1$ was the lattice $A_4^\star$ made of unit hypersheres,
we have thus employed the lattice $A_4^\star$ with covering radius equal exactly to 1.
As consequence of squeezing the resulting grids $S_1$
are made of unit hypersheres and have covering radius less than 1.
We will introduce a one-parameter family of grids
containing the grids $S_1$ constructed for $\cmin<\cmin^*$
(or, equivalently, for $\Delta\omega_0'<\sqrt{2}$)
as a special case and fulfilling the constraints \eqref{wiaz1} and \eqref{wiaz2}.
We expect that for some value of the parameter
we will get grid with covering radius equal exactly to 1,
i.e.\ we will obtain grid with thickness smaller than the thickness of grid $S_1$.
We have found that the construction described below gives grids better than $S_1$
not only when $\Delta\omega_0'<\sqrt{2}$
but also for some values of $\Delta\omega_0'$ greater than and close to $\sqrt{2}$.

We start our construction from replacing the generating matrix $\mathsf{N}$ of the lattice $A_{4}^{\star}$
[given in Eq.\ \eqref{N}] by the matrix built up from the following vectors:
\be
\label{oo}
\begin{array}{l}
\mathbf{o}_1 \coloneqq \mathbf{n}_{1}-\mathbf{n}_{2}-\mathbf{n}_{3}-\mathbf{n}_{4},
\\[1ex]
\mathbf{o}_a \coloneqq \mathbf{n}_a,\quad a=2,3,4.
\end{array}
\ee
The matrix $\mathsf{O}$ with rows made of the components of the vectors $\mathbf{o}_a$
generates the lattice $A_{4}^{\star}$ and has the form
\be
\label{O}
\mathsf{O} = \left(
\begin{array}{cccc}
\frac{1}{2\sqrt{2}}
& -\frac{\sqrt{\frac{5}{6}}}{2}
& -\frac{\sqrt{\frac{5}{3}}}{2}
& -\frac{\sqrt{5}}{2}
\\
\frac{1}{2\sqrt{2}}
& \frac{\sqrt{\frac{15}{2}}}{2}
& 0
& 0
\\
\frac{1}{2\sqrt{2}}
& -\frac{\sqrt{\frac{5}{6}}}{2}
& \sqrt{\frac{5}{3}}
& 0
\\
\frac{1}{2\sqrt{2}}
& -\frac{\sqrt{\frac{5}{6}}}{2}
& -\frac{\sqrt{\frac{5}{3}}}{2}
& \frac{\sqrt{5}}{2}
\end{array}\right).
\ee
Let us make the tails of the vectors \eqref{oo} to coincide with the origin of the coordinate system,
then one can view on them as side edges of some simplex with the height determined by the vector
\be
\label{h2}
\mathbf{h}_{2} \coloneqq \frac{1}{4}\sum_{a=1}^{4}\,\mathbf{o}_a
= \left(\frac{1}{2 \sqrt{2}},0,0,0\right)^\mathsf{T}.
\ee
Euclidean lengthes of the vectors $\mathbf{o}_a$ ($a=1,2,3,4$) and the angles between any two of them
are the same as for vectors $\mathbf{m}_a$ ($a=1,2,3,4$) introduced in \eqref{m1}
[see Eqs.\ \eqref{m3} and \eqref{m4}].
Let us introduce four vectors $\mathbf{a}_a$ perpendicular to the vector $\mathbf{h}_{2}$,
\be
\label{ai}
\mathbf{a}_a := \mathbf{o}_a - \mathbf{h}_2,
\quad a=1,2,3,4.
\ee
One easily checks that $\mathbf{a}_a\cdot\mathbf{h}_2=0$ (for $a=1,2,3,4$).
We also introduce unit vectors
\be
\label{ain}
\widehat{\mathbf{h}}_{2} \coloneqq \frac{\mathbf{h}_{2}}{|\mathbf{h}_{2}|},
\quad
\widehat{\mathbf{a}}_a \coloneqq \frac{\mathbf{a}_a}{|\mathbf{a}_a|},\quad a=1,2,3,4.
\ee

We now \emph{deform} the lattice $A_4^\star$ in the following way:
In the simplex determined by the vectors $\mathbf{o}_a$
we replace the vectors $\mathbf{o}_a$
by the vectors $\mathbf{b}_a$ ($a=1,2,3,4$) of the form
\be
\label{wk}
\mathbf{b}_a(\alpha,\,k) \coloneqq k\cos\alpha\,\widehat{\mathbf{h}}_{2}
+ k\sin\alpha\,\widehat{\mathbf{a}}_a,
\quad a=1,2,3,4,
\ee
where $k>0$ and $0<\alpha<\frac{\pi}{2}$ (so that $0<\cos\alpha<1$) are some parameters.
One easily checks that
\bse
\label{b} 
\begin{align}
|\mathbf{b}_a(\alpha,k)| &= k,\quad a=1,2,3,4,
\\[1ex]
\frac{\mathbf{b}_a(\alpha,k)\cdot\widehat{\mathbf{h}}_2}
{|\mathbf{b}_a(\alpha,k)||\widehat{\mathbf{h}}_{2}|}
&= \cos\alpha, \quad a=1,2,3,4,
\end{align}
\ese
so $k$ is the length of the vectors $\mathbf{b}_a$
and $\alpha$ is the angle between any of the vectors $\mathbf{b}_a$ and the vector $\mathbf{h}_2$.
We are thus changing the lengths of the side edges of the original $A_4^\star$-based simplex
and we are also modifying the angle between the side edges
(and consequently the height of the simplex).
Let us also define the following vector parallel to the $\omega_0'$ axis:
\be
\label{hvec}
\mathbf{h}(\alpha,k) \coloneqq k\cos\alpha\,\widehat{\mathbf{h}}_{2}.
\ee

In the next step we replace one of the vectors $\mathbf{b}_a$, say $\mathbf{b}_{1}$,
by the vector parallel to the $\omega_{0}'$ axis,
this new vector we define as the sum of all four vectors $\mathbf{b}_a$.
We thus get the following tetrad of vectors $\mathbf{c}_a$:
\be
\label{baz}
\begin{array}{l}
\mathbf{c}_{1}(\alpha,k)
\coloneqq \sum_{a=1}^{4}\mathbf{b}_a(\alpha,k)
= 4\mathbf{h}(\alpha,k),
\\[2ex]
\mathbf{c}_{i}(\alpha,k)
\coloneqq \mathbf{b}_a(\alpha,k),\quad a=2,3,4.
\end{array}
\ee
The matrix built up from the vectors $\mathbf{c}_a$ (arranged into its rows) is lower diagonal.
This matrix for $k=\sqrt{2}$ and $\alpha=\arccos(1/4)\cong1.31812$
reproduces the matrix $\mathsf{N}$ from Eq.\ \eqref{N} generating the optimal lattice $A_{4}^{\star}$. 

We require now that the length of the vector $\mathbf{c}_1$
fulfills the constraint
\be
\label{wiaz_k}
|\mathbf{c}_{1}(\alpha,\,k)| = \Delta\omega_{0}',
\ee
where $\Delta\omega_{0}'$ is given in Eq.\ \eqref{omegavsomega}.
Because $|\mathbf{c}_{1}(\alpha,\,k)|=4k\cos\alpha$ [see Eq.\ \eqref{hvec}],
from \eqref{wiaz_k} one can express the length $k$ as a function of the angle $\alpha$: 
\be
\label{k}
k(\alpha) = \frac{\Delta\omega_{0}'}{4\cos\alpha}.
\ee
After substituting \eqref{k} into \eqref{baz},
the generating matrix of the grid built up from the vectors $\mathbf{c}_a$
depends only on the angle $\alpha$ and can symbolically be written as
\be
\label{C2}
\mathsf{C}_{2}(\alpha) = \left(\begin{array}{c}
\mathbf{c}_{1}(\alpha,k(\alpha))^\mathsf{T}
\\[1ex]
\mathbf{c}_{2}(\alpha,k(\alpha))^\mathsf{T}
\\[1ex]
\mathbf{c}_{3}(\alpha,k(\alpha))^\mathsf{T}
\\[1ex]
\mathbf{c}_{4}(\alpha,k(\alpha))^\mathsf{T}
\end{array}\right).
\ee
The matrix $\mathsf{C}_{2}(\alpha)$ is lower diagonal.
Let us note that the matrix $\mathsf{C}_{2}(\alpha)$ reproduces the matrix $\mathsf{C}_1$
(generating the grid $S_1$ in the case $\cmin<\cmin^*$) from Eq.\ \eqref{C1sy}
if one takes $\alpha=\arctan(\sqrt{30}/\Delta\omega_0')$.
The thickness of the grid generated by $\mathsf{C}_{2}(\alpha)$ equals
\be
\rho = \frac{24\sqrt{3}\pi^2\cot^3\alpha}{(\Delta\omega_0')^4}.
\ee

Let us denote by $S_2(\alpha)$ the grid generated by the matrix $\mathsf{C}_{2}(\alpha)$.
We will now find the optimal value of the angle $\alpha$,
i.e.\ this value which minimizes the thickness of the grid $S_2(\alpha)$.
The grid $S_{1}$ was obtained as the result of squeezing the optimal lattice $A_4^\star$
made of unit hypersheres, therefore its covering radius is less than 1.
Let us denote it by $R_{S_{1}}$, $R_{S_{1}}<1$.
We will find the value of the angle $\alpha$
for which the covering radius of the grid
generated by the matrix $\mathsf{C}_{2}(\alpha)$
will be equal to 1.
 
\begin{figure}
\begin{center}
\includegraphics[scale=0.6]{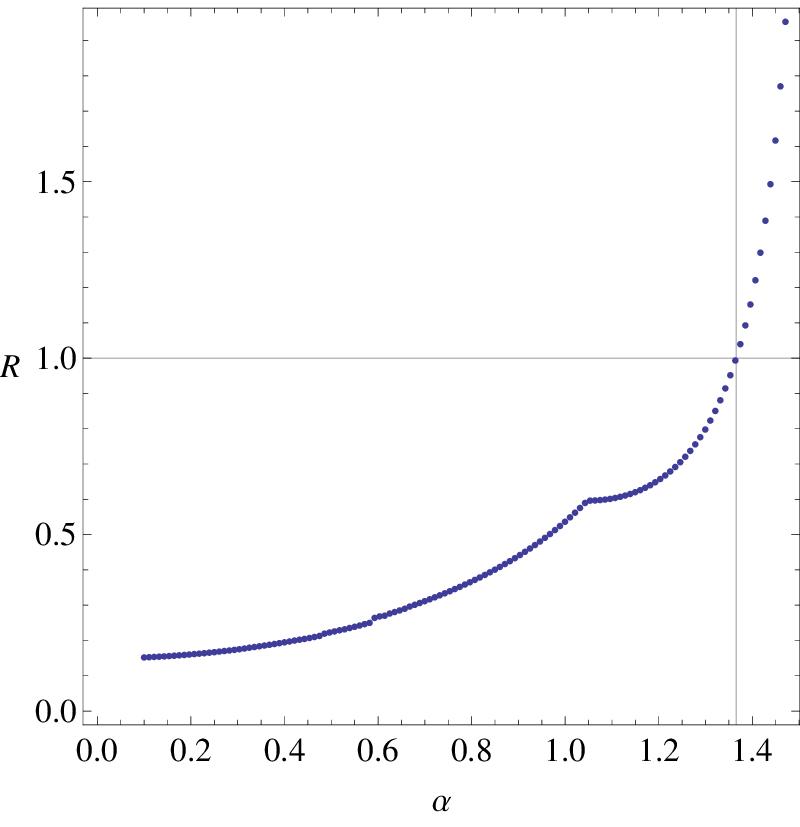}
\caption{\label{r(a)}
Covering radius $R$ of the grid $S_2(\alpha)$ as function of the angle $\alpha$
for $\Delta\omega_{0}'\cong1.19235$.}
\end{center}
\end{figure}

The covering radius of the grid for the given value of the angle $\alpha$
we find by means of the algorithm sketched in Appendix \ref{ACovering}.
This algorithm finds the vertex of the Voronoi cell
inside the fundamental parallelotope spanned by the edge vectors \eqref{wk}.
The covering radius is the distance of this vertex
to the nearest vertex of the fundamental parallelotope.
We have checked that the algorithm described in Appendix \ref{ACovering}
works for angles $\alpha$ in the range $\left(0,\,\frac{\pi}{2}\right)$.

The exemplary dependence of the covering radius of the grid $S_2(\alpha)$ on the angle $\alpha$
is depicted in Fig.\ \ref{r(a)} (for $\Delta\omega_{0}'\cong1.19235$).
We see that the covering radius monotonically increases with the value of the angle $\alpha$.
One can numerically find the value $\alpha_\text{opt}$ of the angle $\alpha$
for which the covering radius is equal to one, $R(\alpha_\text{opt})=1$.
In the case $\Delta\omega_{0}'\cong1.19235$ this value reads
\be
\label{aS2}
\alpha_{\text{opt}} \cong 1.36530894.
\ee
The lengths of the simplex side edges computed [by means of Eq. \eqref{k}]
for this value of the angle $\alpha$ equal $k(\alpha_{\text{opt}})\cong1.46090061$.
The matrix $\mathsf{C}_{2}(\alpha_{\text{opt}})$ generating the best grid $S_2(\alpha_\text{opt})$ reads
\begin{widetext}
\be
\label{S2}
\mathsf{C}_{2}(\alpha_{\text{opt}}) = \left(\begin{array}{cccc}
 1.192353885 & 0 & 0 & 0 \\
 0.298088471 & 1.430165676 & 0 & 0 \\
 0.298088471 & -0.476721892 & 1.348373130 & 0 \\
 0.298088471 & -0.476721892 & -0.674186565 & 1.167725385
\end{array}\right).
\ee
\end{widetext}
The thickness of the grid $S_2(\alpha_\text{opt})$ reads
\be
\label{roS2}
\rho_{S_2(\alpha_\text{opt})}
= \frac{\pi^{2}}{2\,|\det\mathsf{C}_{2}(\alpha_\text{opt})|}
\cong 1.83792.
\ee
It is better by $\sim$12\% than the thickness of the corresponding grid $S_{1}$
(computed for $\Delta\omega_{0}'\cong1.19235$),
\be
\label{procentyS1}
\frac{\rho_{S_{1}}-\rho_{S_2(\alpha_\text{opt})}}{\rho_{S_{1}}} \cong 12.2\%.
\ee
The thickness of the grid $S_{2}$ is only $\sim$4\% larger
than the thickness of the optimal lattice $A_{4}^{\star}$,
\be
\label{procentyA4*}
\frac{\rho_{S_2(\alpha_\text{opt})}-\rho_{A_{4}^{\star}}}{\rho_{A_{4}^{\star}}}
\cong 4.10\%.
\ee

\begin{table}
\caption{\label{table_rho}
Covering thicknesses of the grids $S_1$ and $S_2$
as functions of the minimal value $\cmin$ of the autocovariance
for the three different resolutions $\Delta\omega_0$ of the frequency parameter $\omega_0$:
$\Delta\omega_0\cong4.13044,\ 2.06522,\ 1.03261$.}
\begin{center}
\begin{ruledtabular}
\begin{tabular}{lcccccc}
& \multicolumn{2}{c}{$\Delta\omega_0\cong4.13044$}
& \multicolumn{2}{c}{$\Delta\omega_0\cong2.06522$}
& \multicolumn{2}{c}{$\Delta\omega_0\cong1.03261$}
\\[0.5ex]
\cline{2-3}\cline{4-5}\cline{6-7}
 \multirow{2}{*}{$\cmin$}
&\multirow{2}{*}{$\rho_{S_{1}}$}
&\multirow{2}{*}{$\rho_{S_{2}}$}
&\multirow{2}{*}{$\rho_{S_{1}}$}
&\multirow{2}{*}{$\rho_{S_{2}}$}
&\multirow{2}{*}{$\rho_{S_{1}}$}
&\multirow{2}{*}{$\rho_{S_{2}}$}
\\[1.5ex]\hline\\[-1.5ex]
  0.70 & 1.8135 & 4.5461 & 2.2939 & 1.9174 & 4.5878 & 3.3074 \\
  0.71 & 1.7830 & 4.4753 & 2.2553 & 1.9026 & 4.5107 & 3.2570 \\
  0.72 & 2.0730 & 4.4042 & 2.2161 & 1.8868 & 4.4322 & 3.2056 \\
  0.73 & 2.0356 & 4.3329 & 2.1762 & 1.8700 & 4.3524 & 3.1532 \\
  0.74 & 1.9976 & 4.2612 & 2.1355 & 1.8537 & 4.2710 & 3.1004 \\
  0.75 & 1.9588 & 4.1898 & 2.0940 & 1.8379 & 4.1881 & 3.0466 \\
  0.76 & 1.9192 & 4.1177 & 2.0517 & 1.8229 & 4.1035 & 2.9918 \\
  0.77 & 1.8788 & 4.0459 & 2.0085 & 1.8089 & 4.0171 & 2.9361 \\
  0.78 & 1.8375 & 3.9740 & 1.9644 & 1.7961 & 3.9288 & 2.8792 \\
  0.79 & 1.7953 & 3.9019 & 1.9192 & 1.7848 & 3.8384 & 2.8212 \\
  0.80 & 1.8730 & 3.8303 & 1.8730 & 1.7755 & 3.7459 & 2.7626 \\
  0.81 & 1.8255 & 3.7587 & 1.8255 & 1.7688 & 3.6511 & 2.7009 \\
  0.82 & 1.7769 & 3.6879 & 1.7769 & 1.7657 & 3.5537 & 2.6349 \\
  0.83 & 1.9306 & 3.6175 & 2.1149 & 1.7665 & 3.4536 & 2.5692 \\
  0.84 & 1.8730 & 3.5486 & 2.0517 & 1.7718 & 3.3505 & 2.5066 \\
  0.85 & 1.8135 & 3.4808 & 1.9866 & 1.7922 & 3.2441 & 2.4331 \\
  0.86 & 1.9192 & 3.4157 & 1.9192 & 1.8230 & 3.1341 & 2.3713 \\
  0.87 & 1.8494 & 3.3535 & 1.8494 & 1.8752 & 3.0201 & 2.3041 \\
  0.88 & 1.7769 & 3.2963 & 1.7769 & 1.9638 & 2.9016 & 2.2450 \\
  0.89 & 1.7707 & 3.2457 & 2.1962 & 2.1226 & 2.7781 & 2.1741 \\
  0.90 & 1.8135 & 3.2054 & 2.0940 & 2.4537 & 2.6488 & 2.0989 \\
  0.91 & 1.8315 & 3.1811 & 1.9866 & 3.7346 & 2.5128 & 2.0295 \\
  0.92 & 1.7769 & 3.1342 & 1.8730 & 4.6874 & 2.3691 & 1.9571 \\
  0.93 & 1.8375 & 3.2244 & 2.0730 & 4.4042 & 2.2161 & 1.8868 \\
  0.94 & 1.8135 & 2.8238 & 1.9192 & 4.1178 & 2.0517 & 1.8229 \\
  0.95 & 1.8135 & 3.1055 & 1.8730 & 3.8302 & 1.8730 & 1.7755 \\
  0.96 & 1.8014 & 3.5611 & 1.8730 & 3.5485 & 2.0517 & 1.7718 \\
  0.97 & 1.7769 & 6.7260 & 1.7769 & 3.2962 & 1.7769 & 1.9638 \\
  0.98 & 1.7769 &   & 1.7769 & 3.1342 & 1.8730 & 4.6873 \\
  0.99 & 1.7707 &   & 1.8014 & 3.5613 & 1.8730 & 3.5485 \\
  0.991 & 1.7657 &   & 1.7769 & 4.2229 & 1.7769 & 3.4414 \\
  0.992 & 1.7670 &   & 1.7769 & 5.5767 & 1.8351 & 3.3417 \\
  0.993 & 1.7738 &   & 1.7867 & 9.1866 & 1.7867 & 3.2551 \\
  0.994 & 1.7657 &   & 1.7769 & 39.096 & 1.7769 & 3.1935 \\
  0.995 & 1.7676 &   & 1.7769 &   & 1.7769 & 3.1342 \\
  0.996 & 1.7694 &   & 1.7769 &   & 1.7965 & 2.8233 \\
  0.997 & 1.7676 &   & 1.7916 &   & 1.7769 & 3.2085 \\
  0.998 & 1.7670 &   & 1.7670 &   & 1.7769 & 5.5767 \\
  0.999 & 1.7657 &   & 1.7694 &   & 1.7769 &
\end{tabular}
\end{ruledtabular}
\end{center}
\end{table}

\section{Discussion}

\begin{figure}
\begin{center}
\begin{tabular}{c}
\includegraphics[scale=0.5]{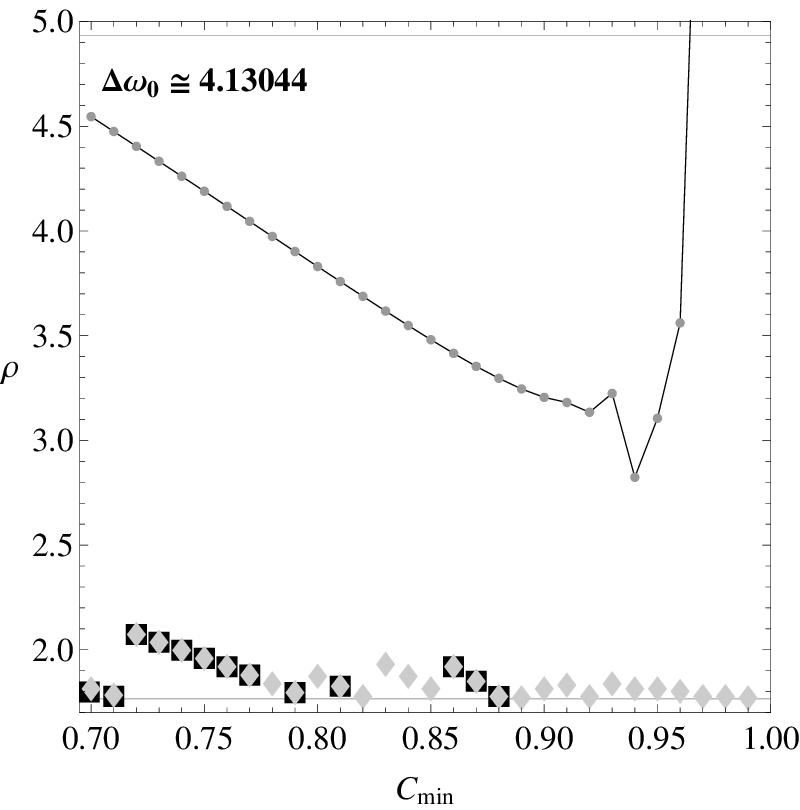}\\
\includegraphics[scale=0.5]{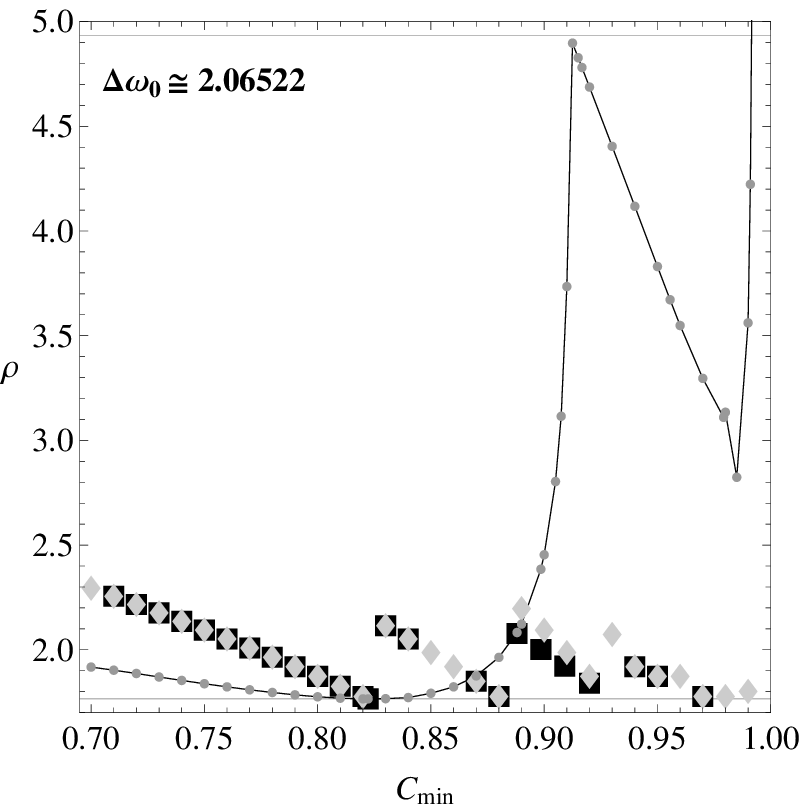}\\
\includegraphics[scale=0.5]{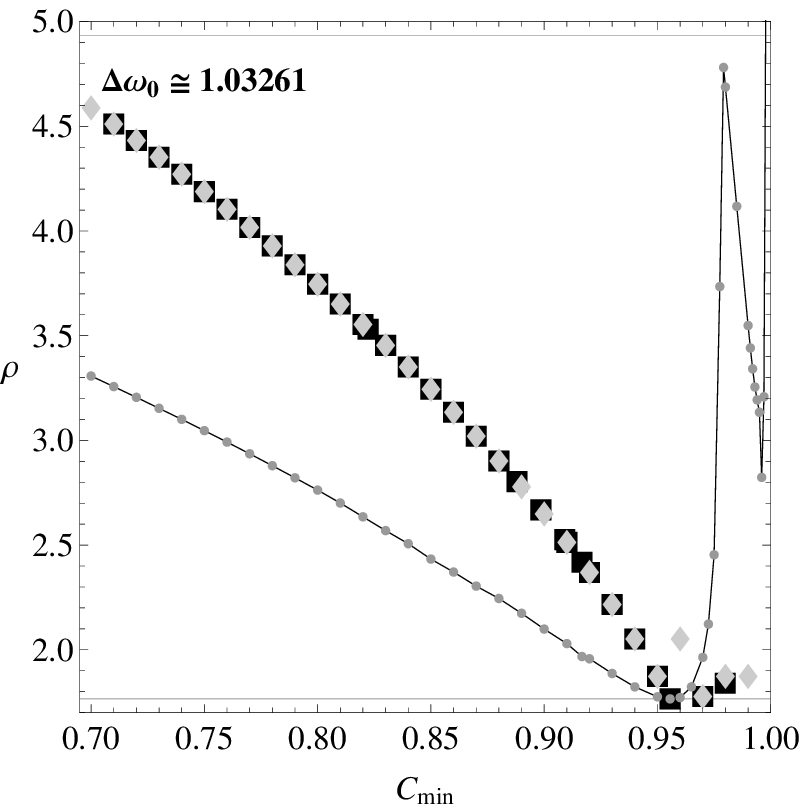}
\end{tabular}
\caption{\label{rho_vs_cmin}
Covering thicknesses $\rho$ of the grids $S_1$ (denoted by diamonds), $S_2$ (circles),
and the grids generated by means of the algorithm taken from Ref.\ \cite{ABJPK10} (squares)
as functions of $\cmin$ for the three different frequency resolutions $\Delta\omega_{0}$ of the search.
The data for the grids $S_1$ and $S_2$ are taken from Table \ref{table_rho}.
The upper horizontal line corresponds to the thickness of the 4-dimensional hypercubic lattice (it equals $\cong4.9348$)
and the lower horizontal line denotes the thickness of the optimal lattice $A_4^\star$ (it equals $\cong1.7655$).}
\end{center}
\end{figure}

We have constructed grids $S_1$ and $S_2$ and computed their thicknesses
for different minimal values $\cmin$ of the autocovariance function
of the $\F$-statistic.
We have taken the values of $\cmin$ from the interval $\left<0.70,\,0.999\right>$
and we have made computations for the three different resolutions $\Delta\omega_0$
of the dimensionless frequency parameter $\omega_0$:
$\Delta\omega_0\cong4.13044,\ 2.06522,\ 1.03261$
[see Eqs.\ \eqref{no.of.data.points}--\eqref{wiaz2n} and the text around them].
The thicknesses of the constructed grids
are listed in Table \ref{table_rho} and are depicted in Fig.\ \ref{rho_vs_cmin}.
Some slots in Table \ref{table_rho} which correspond to grids $S_2$
and high values of $\cmin$ are empty---this is so because the construction
of grids $S_2$ described in Sec.\ \ref{S2grids} works,
for the fixed value of $\Delta\omega_0$,
only up to some maximal value of $\cmin$.

In Fig.\ \ref{rho_vs_cmin}
we have additionally indicated the thicknesses of the grids computed by means of the algorithm
presented in Sec.\ IV of Ref.\ \cite{ABJPK10}.
Not for all combinations of $\Delta\omega_0$ and $\cmin$
we were able to construct grid using this algorithm;
for some of them the algorithm gave no result
after a very long time (of the order of several hours) or just broke down.
For the combinations $(\Delta\omega_0,\cmin)$ for which we have constructed such grids,
in almost all cases they have thickness equal to a good accuracy to the thickness
of the corresponding $S_1$ grid constructed by us.
This is by no means an obvious result,
because the algorithm described in Sec.\ IV of Ref.\ \cite{ABJPK10}
is different from the algorithms devised in the present paper.
We have in particular checked that the basis vectors of grids $S_1$
constructed by us are different from the corresponding basis vectors
produced by the algorithm taken from Ref.\ \cite{ABJPK10}
(obviously with the exception of the first basis vector
which is fixed by the constraint imposed on the grids).
We can thus conclude that the family of grids $S_1$ devised by us
is equivalent (in the sense of possessing the same thickness)
to grids of Ref.\ \cite{ABJPK10} for these values of $(\Delta\omega_0,\cmin)$
for which the algorithm of Ref.\ \cite{ABJPK10} works,
and simultaneously it provides grids for those combinations of $(\Delta\omega_0,\cmin)$
for which the algorithm of \cite{ABJPK10} is unable to produce grids.

The reason of introducing, besides grids $S_1$, the family of grids $S_2$ is
that for some values of search parameters grids $S_2$ have smaller thicknesses than grids $S_1$.
From Fig.\ \ref{rho_vs_cmin} we see that the ranges of the parameter $\cmin$
for which grids $S_2$ have smaller thicknesses than grids $S_1$
are different for different frequency resolutions $\Delta\omega_0$.
For the largest resolution we consider, $\Delta\omega_0\cong4.13044$,
the grids $S_1$ have thicknesses smaller 
than the grids $S_2$ in the whole range $0.70\le\cmin\le0.999$,
so for this value of $\Delta\omega_0$ there is no advantage in using grids $S_2$.
In the worst considered case, which occurs for $\cmin=0.72$, the thickness of the grid $S_1$ is 2.0730
(and it is $\sim$17\% larger than the thickness of the optimal $A_4^\star$ lattice).
For the resolutions $\Delta\omega_0\cong2.06522$ and $1.03261$
the grids $S_2$ are in general better for smaller values of $\cmin$.
In the case $\Delta\omega_0\cong2.06522$ the worst case
(taking into account, for the fixed value of $\cmin$, only the better grid out of $S_1$ and $S_2$ grids)
is for $\cmin=0.89$, then the thickness of the grid $S_2$ is 2.1226, 
$\sim$20\% more than the thickness of the $A_4^\star$ lattice.
For the smallest considered resolution $\Delta\omega_0\cong1.03261$ 
the thicker $S_2$ grids correspond to smaller values of $\cmin$;
the thickest one, for $\cmin=0.70$, has the thickness 3.3074,
$\sim$87\% larger than the thickness of the $A_4^\star$ lattice.

Reference \cite{2014--LSC-VC--CQG-b} reported the results of an $\F$-statistic-based
all-sky search for continuous gravitational waves in Virgo VSR1 data.
In this search the grid computed by means of the algorithm of Ref.\ \cite{ABJPK10}
was employed. The search parameters was: $N=344656$, $N_\text{FFT}=2^{20}$, and $\cmin=0.75$,
what leads to resolutions $\Delta\omega_0\cong2.06522$ and $\Delta\omega_0'\cong1.19235$.
For these values of search parameters our grid $S_1$ has the thickness $\rho_{S_1}\cong2.0940$
[see Eq.\ \eqref{ro_s1}], which is identical with the thickness of the grid based of the algorithm of Ref.\ \cite{ABJPK10}
[see the middle panel of Fig.\ \ref{rho_vs_cmin} for $\cmin=0.75$],
whereas the grid $S_2$ has thickness $\sim$12\% smaller,
$\rho_{S_2}\cong1.8379$ [see Eqs.\ \eqref{roS2}--\eqref{procentyS1}].

\begin{figure*}
\begin{center}
\includegraphics[scale=0.6]{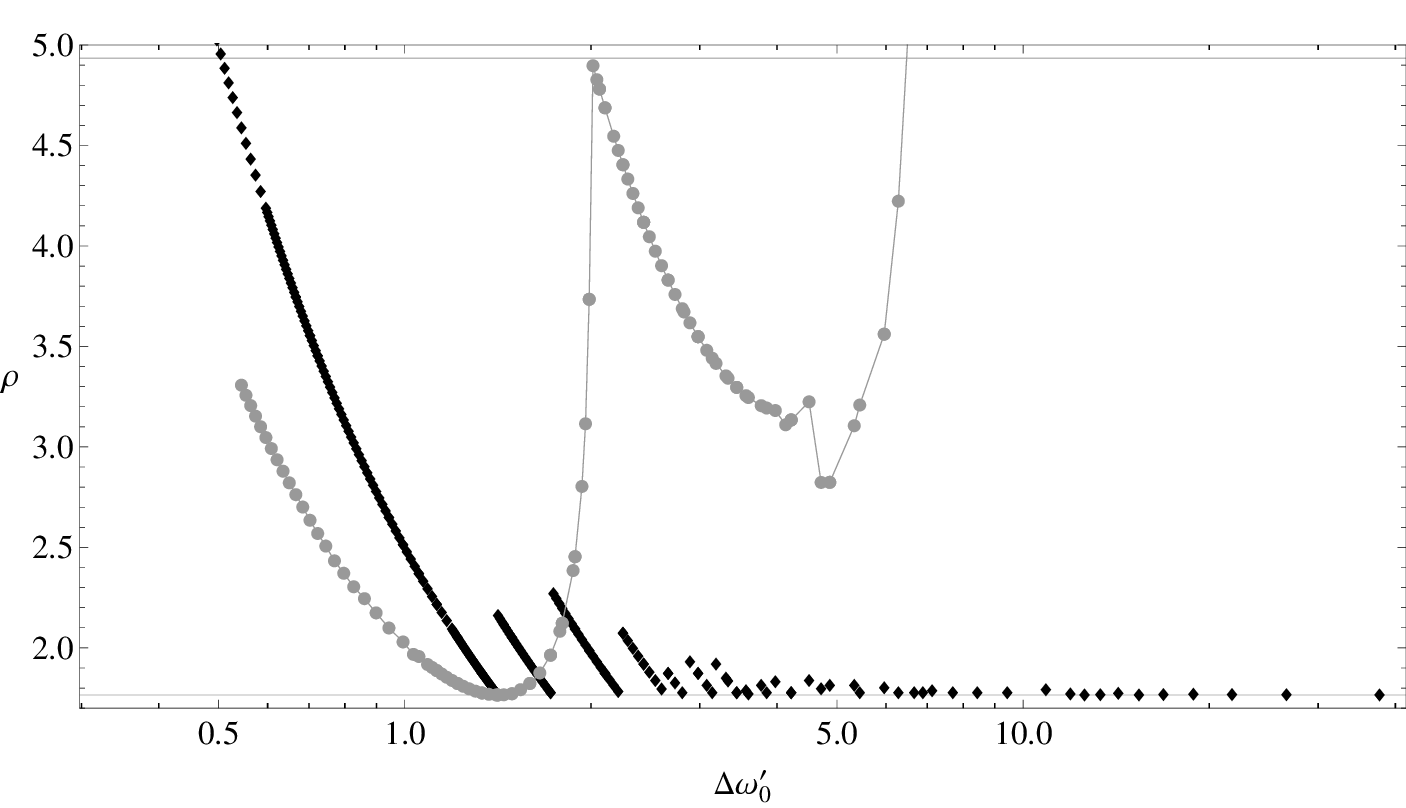}
\caption{\label{s1s2omega1}
Covering thickness $\rho$
of the grids $S_1$ (diamonds) and $S_2$ (circles)
as functions of the quantity $\Delta\omega_0'$.
We have shown here all grids $S_1$ and $S_2$
depicted in three panels of Fig.\ \ref{rho_vs_cmin}
as well as some more grids computed for smaller values of $\Delta\omega_0'$.
The upper horizontal line corresponds to the thickness of the 4-dimensional hypercubic lattice (it equals $\cong4.9348$)
and the lower horizontal line denotes the thickness of the optimal lattice $A_4^\star$ (it equals $\cong1.7655$).}
\end{center}
\end{figure*}

All grid constructions devised in Sec.\ IV of our paper
depend on  the search parameters $\Delta\omega_0$ and $\cmin$
only through the quantity $\Delta\omega_0'$, see Eq.\ \eqref{omegavsomega}.
It means that any two grids
(belonging to $S_1$ or to $S_2$ family)
depicted on different panels of Fig.\ \ref{rho_vs_cmin}
which are determined by search parameters
$(\Delta\omega_0^1,\cmin^1)$ and $(\Delta\omega_0^2,\cmin^2)$
such that $\Delta\omega_0^1/\sqrt{1-\cmin^1}=\Delta\omega_0^2/\sqrt{1-\cmin^2}$,
leads to the same value of $\Delta\omega_0'$ [see Eq.\ \eqref{omegavsomega}],
they thus both correspond to the same grid in $\Omega'$ space.
In Fig.\ \ref{s1s2omega1} we have shown
the dependence of the thicknesses of the grids $S_1$ and $S_2$
on the value of the parameter $\Delta\omega_0'$.
One can see that the thicknesses of the grids $S_1$ depicted in Fig.\ \ref{s1s2omega1}
split into several branches (better visible for smaller values of $\Delta\omega_0'$).
These branches correspond to different
choices of the lattice vector $\mathbf{q}$ introduced in Sec.\ IV~C.
According to Eq.\ \eqref{rhoS1b2} the thickness of the grid $S_1$
is proportional to the length of $\mathbf{q}$, so the sequence
of branches visible in Fig.\ \ref{s1s2omega1} is in fact
the sequence of the lengths of possible lattice vectors
for the optimal lattice $A_4^\star$
[the squares of these lengths are given in Eq.\ \eqref{A4lengths}].
The first branch (going from left to right) corresponds to $|\mathbf{q}|=\sqrt{2}$
[see Eq.\ \eqref{A4lengths}], the second one to $|\mathbf{q}|=\sqrt{3}$, and so on.

\begin{acknowledgments}

The work presented in this paper was supported in part
by the Polish Ministry of Science and Higher Education
grants no.\ N N203 387237 and DPN/N176/VIRGO/2009.
The contribution of A.~Pisarski was also supported
by the Project \emph{Podlaski Fundusz Stypendialny}
of the Operational Programme Human Capital
(Priority VIII, Regional Human Resources of the Economy).
We would like to thank Andrzej Kr\'olak for helpful discussions.

\end{acknowledgments}

\appendix

\section{Comparison with the language of a~metric in the space of signal's parameters}
\label{Ametric}

In construction of template banks for searches of gravitational waves
(not necessarily of continuous type)
one can employ a geometric approach based on a notion of a \emph{metric} introduced in the space of signal's parameters.
We follow here derivation of this metric given in Sec.\ 2 of Ref.\ \cite{P07}.
In the general case of template-based searches one constructs a \emph{detection statistic} $\F[x;\btheta]$,
which depends on data $x$ and on the parameters $\btheta$ of the signal we are looking for.
Then one considers the expectation value $\bar{\F}(\btheta,\btheta'):=\mathrm{E}_1\{\F[x;\btheta]\}$
of the detection statistic in the case when data contains the signal with the parameters $\btheta'$
[i.e., when $x(t)=n(t)+h(t;\btheta')$].
One assumes that the expectation value $\bar{\F}(\btheta,\btheta')$
has a maximum at $\btheta=\btheta'$, so
\be
\frac{\partial\bar{\F}(\btheta,\btheta')}{\partial\theta^i}\bigg|_{\btheta=\btheta'} = 0.
\ee
One can define a \emph{mismatch} $m$ which characterizes the fractional loss
in the expected value of the detection statistic $\F$,
\be
\label{mismatch}
m(\btheta,\btheta') := 1 - \frac{\bar{\F}(\btheta,\btheta')}{\bar{\F}(\btheta',\btheta')}.
\ee
The quantity
\be
\label{match}
M(\btheta,\btheta') := 1 - m(\btheta,\btheta')
= \frac{\bar{\F}(\btheta,\btheta')}{\bar{\F}(\btheta',\btheta')}
\ee
can thus be called a \emph{match}.

One expands the right-hand side of the definition \eqref{mismatch}
with respect to small parameter \emph{offsets} $\Delta\btheta:=\btheta-\btheta'$,
\be
\label{metric1}
m(\btheta'+\Delta\btheta,\btheta') = \sum_{i,j} g_{ij}(\btheta') \Delta\theta^i\Delta\theta^j
+ \mathcal{O}(\Delta\theta^3),
\ee
where $g_{ij}(\btheta')$ is the positive-definite metric tensor on the parameter space,
\be
\label{metric2}
g_{ij}(\btheta') := -\frac{1}{2\bar{\F}(\btheta',\btheta')}
\frac{\partial^2 \bar{\F}(\btheta,\btheta')}{\partial\theta^i\partial\theta^j}\bigg|_{\btheta=\btheta'}.
\ee

In constructions of template banks needed for all-sky searches for continuous signals considered in our paper
it is natural to take the $\F$-statistic [introduced in Eq.\ \eqref{29}] as a detection statistic.
Its expectation value $\bar{\F}(\btheta,\btheta')$ is given by Eq.\ \eqref{EF3},
\be
\label{E1F}
\bar{\F}(\btheta,\btheta')\cong 1 + \frac{1}{2} \, \rho(h_0')^2 \, C_0(\bxi,\bxi'),
\ee
where $\rho(h_0')$ is the optimal signal-to-noise ratio [given in Eq.\ \eqref{snr}].
The mismatch \eqref{mismatch} is then equal
\be
m(\btheta,\btheta') \cong 1 - \frac{1 + \frac{1}{2}\,\rho(h_0')^2\,C_0(\bxi,\bxi')}{1 + \frac{1}{2}\,\rho(h_0')^2 }.
\ee
For large signal-to-noise ratios, $\rho(h_0')\gg1$, this can be approximate further,
\be
\label{mC0}
m(\bxi,\bxi') \cong 1 - C_0(\bxi,\bxi').
\ee
This is the relation between the mismatch and the autocovariance function
(we have replaced here the arguments $\btheta,\btheta'$ of $m$ by $\bxi,\bxi'$,
as now the mismatch depends only on intrinsic parameters of the signal).
The autocovariance function $C_0(\bxi,\bxi')$ plays thus the role of the match introduced in Eq.\ \eqref{match},
the quantity $\cmin$ we can identify with the \emph{minimal match}
and $1-\cmin$ with the \emph{maximal mismatch} of a template bank.
Be aware that sometimes instead of $\cmin$ the quantity $\sqrt{\cmin}$ is called the minimal match
(this is so e.g.\ in the monograph \cite{JKbook}, see Eq.\ (7.32) there,
and also in \cite{ABJPK10}).

For the phase of the gravitational-wave signal considered by us
[it has the form given in Eq.\ \eqref{Phi}],
the autocovariance $C_0(\bxi,\bxi')$ is a function of $\btau:=\bxi-\bxi'$
and for $|\btau|\ll1$ it can be approximated by Eq.\ \eqref{Ca},
what leads to equality
\be
m(\bxi'+\btau,\bxi') \cong \sum_{k,l} \tilde{\Gamma}_{kl}\,\tau_k\,\tau_l.
\ee
By comparing this with Eq.\ \eqref{metric1} we see that the reduced Fisher matrix $\tilde{\Gamma}$
plays the role of a metric in the space of intrinsic parameters of the signal.

Let us finally note the the equation \eqref{mC0} can also be interpreted in another way.
Following Jaranowski and Kr\'olak [see Eqs.\ (103)--(105) in Ref.\ \cite{JK00}]
and Prix [see Eq.\ (28) in Ref.\ \cite{Prix2007}],
one can define suboptimal (or mismatched)
signal-to-noise ratio which can be achieved by using the template with parameters $\btheta$
to detect a signal with parameters $\btheta'$,
\be
\rho_\text{sub} := \sqrt{\mathrm{E}_1\{2\F[x;\btheta]\} - 2}.
\ee
By virtue of \eqref{E1F} we have
\be
\rho_\text{sub}(h_0',\bxi,\bxi') \cong \rho(h_0')\,\sqrt{C_0(\bxi,\bxi')},
\ee
the mismatch \eqref{mC0} can thus be interpreted as a fractional loss of the (squared)
signal-to-noise ratio,
\be
m(\bxi,\bxi') \cong 1 - \frac{\rho_\text{sub}(h_0',\bxi,\bxi')^2}{\rho(h_0')^2}.
\ee

\section{Lengths of the lattice vectors for the optimal lattice $A_4^\star$}
\label{ALengths}

Let the origin of the coordinate system coincides with some node of the $A_4^\star$ lattice,
then one can compute the lengths of the vectors joining the origin with other nodes of the lattice.
We have checked that the first 119 smallest \emph{squares} of the lengths
can be written as
\be
\label{A4lengths}
10n,\; 10n+2,\;10n+3,\;10n+5,\;10n+7,\;10n+8,
\ee
where $n\in\{0,1,\ldots,19\}$, so the lengths form the following increasing sequence
\begin{multline}
2, 3, 5, 7, 8, 10, 12, 13, 15, 17, 18,  \ldots
\\
\ldots, 190, 192, 193, 195, 197, 198.
\end{multline}

\begin{figure*}
\begin{center}
\begin{tabular}{c}
\includegraphics[scale=.5]{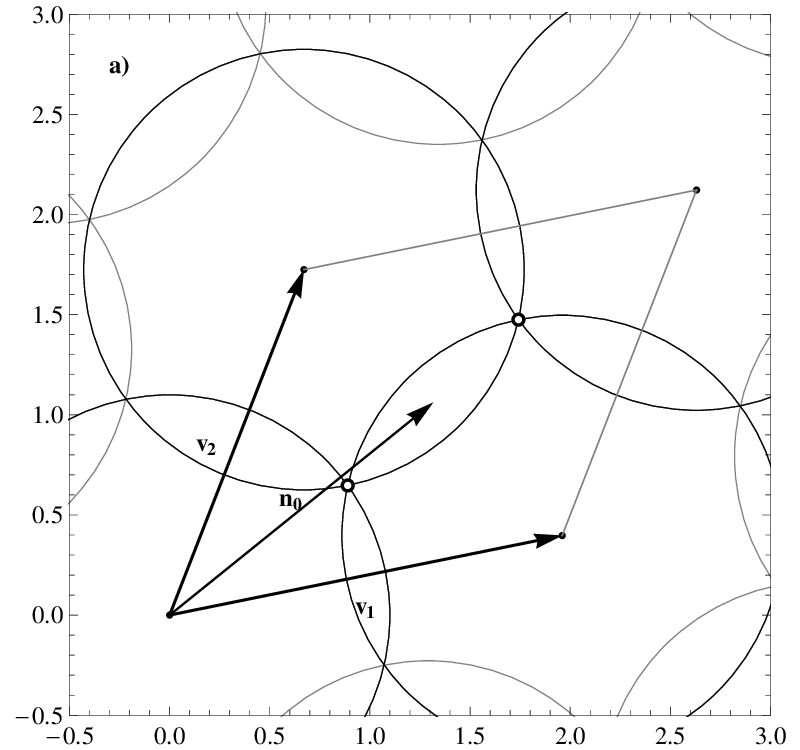}
\includegraphics[scale=.5]{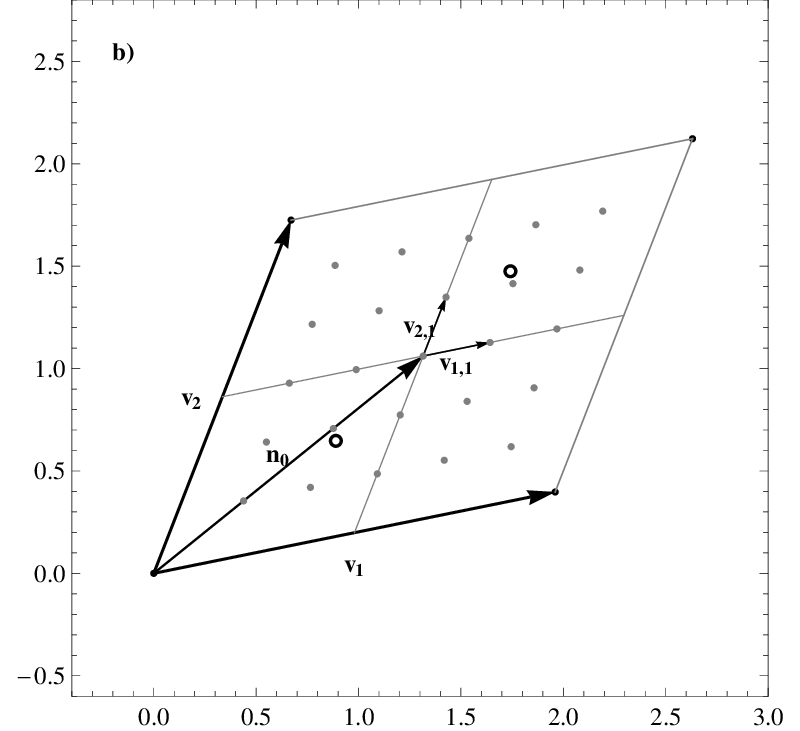}\\
\includegraphics[scale=.5]{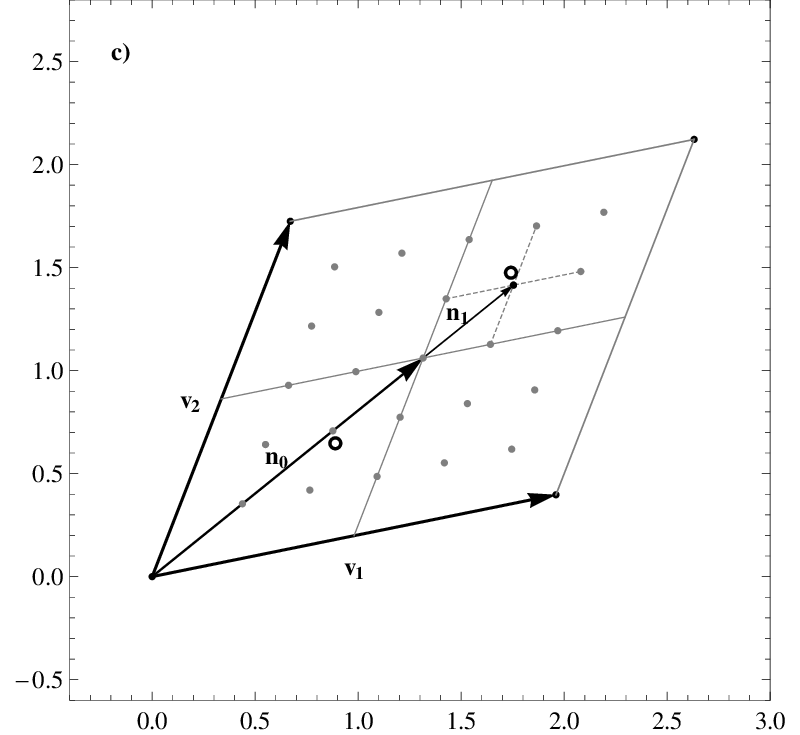}
\includegraphics[scale=.5]{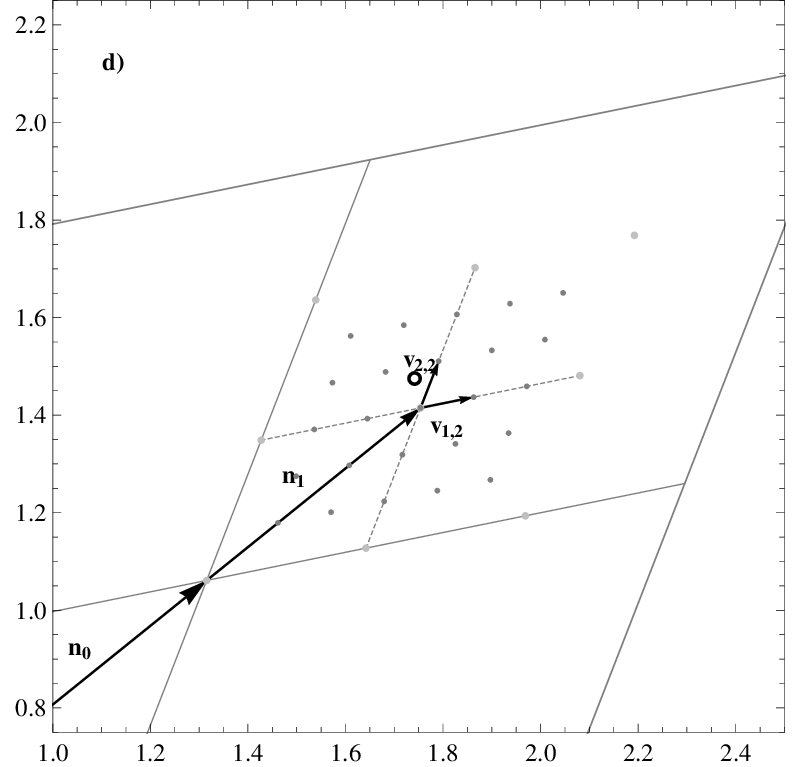}\\
\includegraphics[scale=.5]{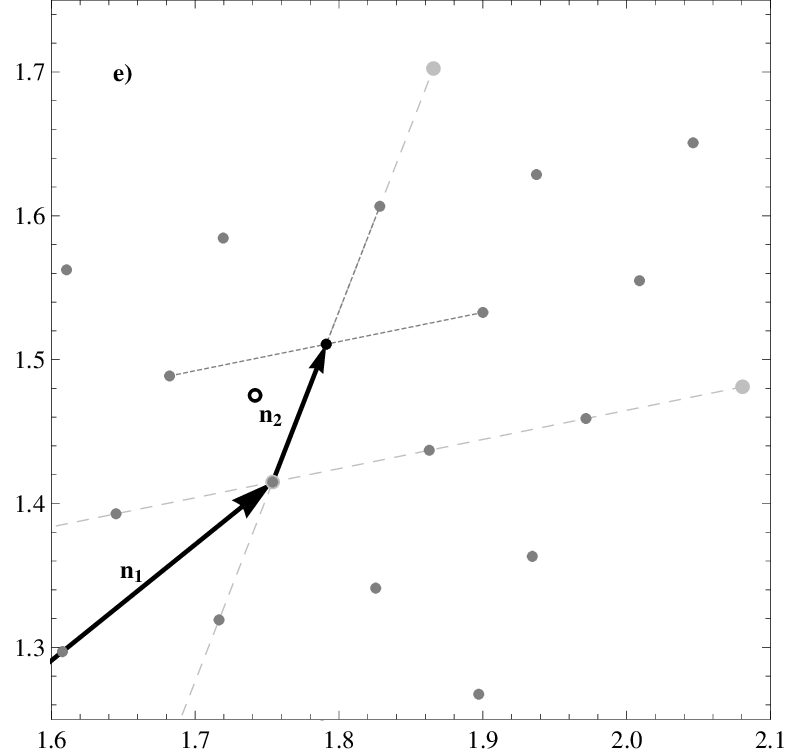}
\includegraphics[scale=.5]{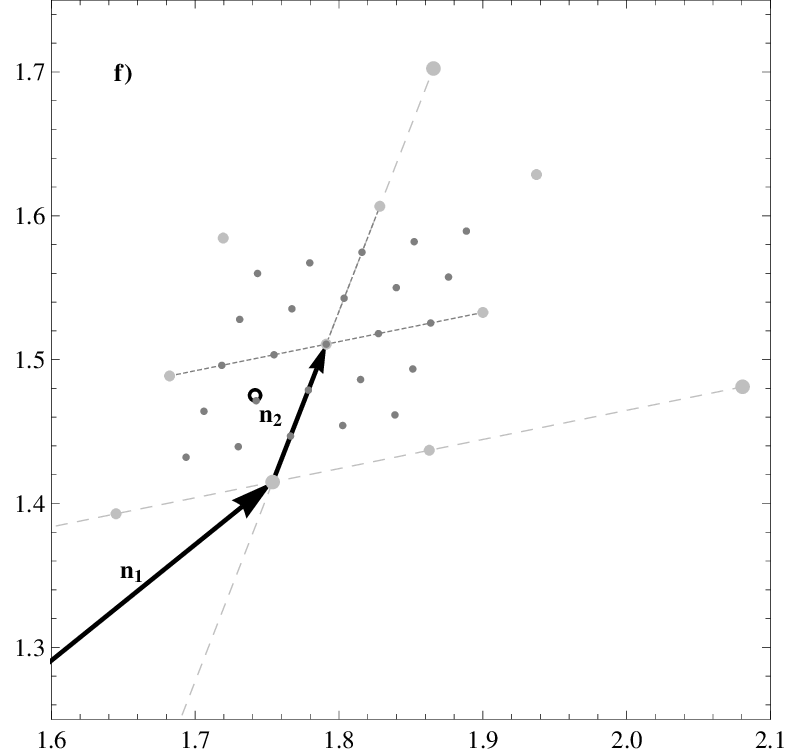}\\
\end{tabular}
\caption{\label{Afigure}
Algorithm finding covering radius of given lattice.
The vertexes of the Voronoi cell are denoted by open circles.
a) The vector $\mathbf{n}_0$ coincides with the center of the fundamental parallelotope.
b) Around the head of the vector $\mathbf{n}_0$ we construct grid of points determined by Eq.\ \eqref{Awektora}.
For each point we compute the minimal distance of it to the vertexes of the fundamental parallelotope.
In c) we pick up the point $\mathbf{n}_1$ for which this minimal distance is maximal. 
Images d)--f) illustrates the next stages of the algorithm.}
\end{center}
\end{figure*}

\section{Algorithm finding covering radius}
\label{ACovering}

The algorithm searches within fundamental parallelotope of given lattice vertexes of its Voronoi cell,
i.e.\ the points which are the most distant from any vertex of the fundamental parallelotope.
The algorithm makes use of the function, which for a given set of points lying inside the fundamental
parallelotope, computes the minimal distance of every point from all vertexes of the parallelotope,
and picks up this point for which the minimal distance achieves maximum.

The algorithms works iteratively.
Let the fundamental parallelotope of some $d$-dimensional lattice be spanned
by the vectors $(\mathbf{v}_{1},\dots,\mathbf{v}_d)$
and let $\mathbf{n}_{i}$ ($i=1,2,\ldots$) be the position of the point picked up at the $i$th stage.
The the points considered in the $(i+1)$th stage are determined by the formula
\be
\label{Awektora}
\mathbf{n}_i = \mathbf{n}_{i-1} + \sum_{m=1}^d a_m \mathbf{v}_{m,i},
\quad a_m \in \{-2,1,0,1,2\},
\ee
where
\be
\mathbf{v}_{m,i} \coloneqq  \frac{1}{6}\frac{1}{3^{i-1}} \mathbf{v}_m,
\quad i=1,2,\ldots\ .
\ee
The vector $\mathbf{n}_0$ initializing the algorithm
coincides with the geometrical center of the fundamental parallelotope.
Operation of the algorithm is illustrated in Fig.\ \ref{Afigure}
for some 2-dimensional lattice.

\end{document}